# The Covariant Approach to LRS Perfect Fluid Spacetime Geometries

Henk van Elst[1]*& George F R Ellis[2,1†]

1 *School of Mathematical Sciences, Queen Mary & Westfield College, Mile End Road London E1 4NS, United Kingdom*

2 *Department of Applied Mathematics, University of Cape Town, Rondebosch 7700 Cape Town, South Africa*

**Abstract**

The dynamics of perfect fluid spacetime geometries which exhibit *Local Rotational Symmetry* (LRS) are reformulated in the language of a $1 + 3$ "threading" decomposition of the spacetime manifold, where covariant fluid and curvature variables are used. This approach presents a neat alternative to the orthonormal frame formalism. The dynamical equations reduce to a set of differential relations between purely scalar quantities. The consistency conditions are worked out in a transparent way. We discuss their various subcases in detail and focus in particular on models with higher symmetries within the class of expanding spatially inhomogeneous LRS models, via a consideration of functional dependencies between the dynamical variables.

PACS number(s): 98.80.Hw, 04.20.-q, 04.20.Jb, 97.60.-s

*e-mail: H.van.Elst@maths.qmw.ac.uk

†e-mail: ELLIS@maths.uct.ac.za

# 1 Introduction

In studying the geometry and dynamics of relativistic cosmological models two main streams have occurred: one [1, 2] focusing on the spacetime (Killing) symmetries of these models, and one [3, 4] focusing on their covariant properties arising from a $1+3$ "threading" decomposition of the spacetime manifold with respect to an invariantly defined normalised timelike congruence. It is commonplace to determine the covariant properties of models when the prime focus of analysis is imposed spacetime symmetries, but the converse case has not been systematically addressed. This paper is one of a series that aims to fill this gap, that is, to determine how spacetime symmetries appear when models have been characterised in terms of their covariant description. There are two main cases to consider here. The first is when there is no continuous isotropy group: there exist at most discrete isotropies in the cosmological model. That will be the subject of a later paper. The second is when they are *Locally Rotationally Symmetric* (LRS) [5, 6], that is there exists a continuous isotropy group at each point, and consequently, a multiply-transitive isometry group acting on the spacetime manifold . Such models are the concern of this paper, which aims to covariantly characterise all cosmological models that are LRS and have a perfect fluid matter source.

LRS spacetime geometries have been discussed many times in the literature, usually in terms of a local coordinate or an orthonormal frame formulation. It is known that in the cosmological context, the isotropies around a point can occur as a 1-D or 3-D subgroup of the full group of isometries, which leaves the normalised 4-velocity $u^i/c$ of the matter fluid flow invariant. The latter case are the everywhere isotropic Friedmann–Lemaître–Robertson–Walker (FLRW) spacetime geometries, that are well understood (and are the standard models of cosmology). The former are anisotropic and in general spatially inhomogeneous (the spatially homogeneous ones are characterised as subcases of the Bianchi models in [7, 8]). We will consider their nature and classification in detail in what follows.

In the covariant $1+3$ "threading" picture the invariantly defined normalised 4-velocity $u^i/c$ (such that $u_i/c\,u^i/c = -1$) of the matter fluid flow and a tensor projecting orthogonal to it, defined by $h^i{}_j := \delta^i{}_j + u^i/c\,u_j/c$, are the key ingredients. We assume that the matter content of the cosmological model at hand, which in the fluid description in general has stress-energy-momentum tensor

$$T_{ij} = p\,h_{ij} + \pi_{ij} + 2\,q_{(i}/c\,u_{j)}/c + \mu\,u_i/c\,u_j/c \tag{1}$$

when decomposed into its irreducible parts with respect to $u^i/c$, has the characteristic features of a perfect fluid (vanishing energy current density, $q^i = 0$, and vanishing anisotropic pressure, $\pi_{ij} = 0$), and that the matter fluid is isentropic, so that its total energy density $\mu$ and its isotropic pressure $p$ are related via a barotropic equation of state $p = p(\mu)$ [9, 10] (with $\frac{\partial p}{\partial \mu} \neq 0$, unless stated otherwise). Through this assumption a lot of relativistic thermodynamical processes of physical interest involving viscous effects and entropy production are excluded. Nevertheless, these simplifications on the side of the matter source provide the possibility of a neat treatment of the geometrical properties of the above mentioned cosmological models with some spatial isotropy. We also exclude the vacuum case, that is, we assume $(\mu + p) > 0$. The usual definitions of the kinematic quantities (see [3, 4]) are employed and we follow the sign and index conventions of Kramer et al [1].

When the spacetime is LRS, there exists a unique preferred spatial direction at each point, covariantly defined for example by either a vorticity vector field, an eigendirection of a rate of shear tensor field, or a non-vanishing non-gravitational acceleration of the matter fluid elements. This direction constitutes a local axis of symmetry: all observations are identical under rotations about it, and in particular they are the same in all (spatial) directions perpendicular to that direction [5, 6]. Thus there exists a preferred spacelike unit vector field $e^i$

$$e_i\,u^i/c = 0\,, \quad e_i\,e^i = 1 \qquad \Rightarrow \qquad e^i\,(\nabla_j e_i) = 0\ . \tag{2}$$

Due to the local rotational symmetry, *all* covariantly defined spacelike *vector fields* in the spacetime must be proportional to $e^i$, that is, on defining as usual ( $\eta_{1234} = -\sqrt{-g}$ )

$$\nabla_i u_j/c := \sigma_{(ij)}/c + \tfrac{1}{3}\,(\Theta/c)\,h_{ij} - \eta_{ijkl}\,\omega^k/c\,u^l/c - u_i/c\,\dot{u}_j/c^2\ , \tag{3}$$

with $\sigma^i{}_i = 0$ and $\omega^i/c := -1/2\,\eta^{ijkl}\,\omega_{jk}/c\,u_l/c \Leftrightarrow \omega_{ij}/c = -\,\eta_{ijkl}\,\omega^k/c\,u^l/c$, we must have

$$\begin{array}{rclrclrcl} \dot{u}^i/c^2 &=& (\dot{u}/c^2)\,e^i\ , & h^j{}_i\,\nabla_j\mu &=& \mu'\,e_i\ , & h^j{}_i\,\nabla_j p &=& p'\,e_i\ , \\ \omega^i/c &=& (\omega/c)\,e^i\ , & h^j{}_i\,\nabla_j\,(\Theta/c) &=& (\Theta/c)'\,e_i\ . && \end{array} \tag{4}$$

This applies equally to the derivatives of $e^i$: using (2),

$$h^i_{\ j}\,(e^j)' \;:=\; h^i_{\ j}\,e^k\,(\nabla_k e^j) = 0\,, \tag{5}$$

$$h^i_{\ j}\,(e^j)^{\cdot}/c \;:=\; h^i_{\ j}\,u^k/c\,(\nabla_k e^j) = 0\,, \tag{6}$$

where a prime denotes the covariant derivative along the preferred direction $e^i$, and a dot the covariant time derivative along the fluid flow lines. Hence, the preferred spacelike direction $e^i$ is geodesic in the local 3-spaces orthogonal to $u^i/c$ and Fermi-propagated along the matter fluid flow lines. The spatial rotation ("twist") of $e^i$ must be proportional to $e^i$ itself, that is

$$\eta^{ijkl}\,(\nabla_j e_k)\,u_l/c = -\,k\,e^i\,, \tag{7}$$

where $k$ denotes the magnitude of the spatial rotation of $e^i$: $k := |\,\eta^{ijkl}\,(\nabla_j e_k)\,u_l/c\,|$. We define $a$ as the magnitude of the spatial divergence of $e^i$:

$$a := h^j_{\ i}\,(\nabla_j e^i) \quad\Rightarrow\quad \nabla_i e^i = a + (\dot{u}/c^2)\ . \tag{8}$$

A unique spacelike *tracefree symmetric tensor field* $e_{ij}$ is defined from $e^i$ by

$$e_{(ij)} := \tfrac{1}{2}\,(3\,e_i\,e_j - h_{ij}) := h_{ij} - \tfrac{3}{2}\,p_{(ij)}\ , \tag{9}$$

which has the properties that $e_{ij}\,u^j = 0$, $e_{ij}\,e^j = e_i$, $e^i_{\ i} = 0$, $e_{ik}\,e^k_{\ j} = 1/2\,(e_{ij} + h_{ij})$ and $e^i_{\ j}\,e^j_{\ i} = 3/2$. We define a tensor projecting orthogonal to both $e^i$ and $u^i/c$ by $p^i_{\ j} := h^i_{\ j} - e^i\,e_j$. The spatial divergence and the covariant time derivative of $e_{ij}$ are

$$h^i_{\ k}\,h^j_{\ l}\,(\nabla_j e^{kl}) \;=\; \tfrac{3}{2}\,h^j_{\ k}\,(\nabla_j e^k)\,e^i = \tfrac{3}{2}\,a\,e^i\ , \tag{10}$$

$$h^k_{\ i}\,h^l_{\ j}\,(e_{kl})^{\cdot}/c \;=\; 0\ . \tag{11}$$

Again, due to the assumed LRS symmetry of the spacetime manifold, *all* tracefree symmetric tensor fields have to be proportional to $e_{ij}$. Thus, with the definitions

$$E_{ij} \;:=\; C_{ikjl}\,u^k/c\,u^l/c \;=\; E_{(ij)}\ , \tag{12}$$

$$H_{ij} \;:=\; -\,\tfrac{1}{2}\,\eta_{iklm}\,C^{lm}_{\ \ jn}\,u^k/c\,u^n/c \;=\; H_{(ij)}\ , \tag{13}$$

which imply $E^i_{\ i} = 0$, $H^i_{\ i} = 0$, we have for the rate of shear tensor and the "electric" and "magnetic parts" of the tracefree Weyl conformal curvature tensor respectively

$$\sigma_{ij} = \tfrac{2}{\sqrt{3}}\,\sigma\,e_{ij}\ , \qquad E_{ij} = \tfrac{2}{\sqrt{3}}\,E\,e_{ij}\ , \qquad H_{ij} = \tfrac{2}{\sqrt{3}}\,H\,e_{ij}\ , \tag{14}$$

where we define the squared magnitudes $\sigma^2 := 1/2\,\sigma^i_{\ j}\,\sigma^j_{\ i} \geq 0$, $E^2 := 1/2\,E^i_{\ j}\,E^j_{\ i} \geq 0$, $H^2 := 1/2\,H^i_{\ j}\,H^j_{\ i} \geq 0$. Note that $\sigma$, $E$ and $H$ as defined in Eq. (14) can be either positive or negative. It follows that all solutions are of Petrov type D if the Weyl curvature tensor is non-zero [11, 1]. Thus we now find that

$$\nabla_i u_j/c = \tfrac{2}{\sqrt{3}}\,(\sigma/c)\,e_{ij} + \tfrac{1}{3}\,(\Theta/c)\,h_{ij} + (\omega/c)\,s_{[ij]} - (\dot{u}/c^2)\,u_i/c\,e_j\ , \tag{15}$$

with the antisymmetric tensor definition $s_{[ij]} := -\,\eta_{ijkl}\,e^k\,u^l/c$. Further, on taking the orthogonality relation $e^i\,(\nabla_j e_i) = 0$ into account, we can express the total spatial projection of the covariant derivative of $e^i$ as

$$h^k_{\ i}\,h^l_{\ j}\,(\nabla_k e_l) = \tfrac{a}{2}\,p_{(ij)} + \tfrac{k}{2}\,s_{[ij]}\ . \tag{16}$$

Expanding out the projection tensors and remembering (2), this leads to

$$\nabla_i e_j = \tfrac{a}{2}\,p_{(ij)} + \tfrac{k}{2}\,s_{[ij]} + \left[\ \tfrac{2}{\sqrt{3}}\,(\sigma/c) + \tfrac{1}{3}\,(\Theta/c)\ \right]\,e_i\,u_j/c - (\dot{u}/c^2)\,u_i/c\,u_j/c\ . \tag{17}$$

From the point of view of the Cartan–Karlhede equivalence problem formalism for invariantly classifying different spacetime geometries [12, 13], in the covariant $1+3$ approach to LRS spacetimes we condense rotational isometries in the introduction of the spacelike unit vector field $e^i$. To cover translational isometries we can specify a maximal set of four generalised essential coordinates, which we then can investigate for the existence of functional dependencies [12]. For (non-vacuum) models with non-vanishing rate of expansion $(\Theta/c)$ we choose this set to be constituted by the four covariantly defined scalars $S_4 := \{\,\mu, (\Theta/c)\,, (\sigma/c)\,, E\,\}$.

LRS perfect fluid spacetime geometries found frequent application in the literature in for example the modeling of the dynamical processes surrounding the formation of relativistic stars or galaxies ([14], [15], [9], [10]). Irrespective of the large degree of idealisation of the physics underlying their evolution, they have proved to be a valuable test ground for more complex astrophysical scenarios.

# 2 The Equations

Using the relations introduced in the previous section, we can express all the covariant equations of Ref. [4] as equations for covariantly defined scalars $f$. These are in our case $\mu$, $p$, $(\Theta/c)$, $(\sigma/c)$, $(\omega/c)$, $\left(\dot{u}/c^2\right)$, $a$, $k$, $E$, and $H$.

## 2.1 The Ricci identities

### 2.1.1 Time derivative equations

$$\begin{aligned}
(\Theta/c)^{\cdot}/c &= -\tfrac{1}{3}\,(\Theta/c)^2 + \left(\dot{u}/c^2\right)' + a\,\left(\dot{u}/c^2\right) + \left(\dot{u}/c^2\right)^2 - 2\,(\sigma/c)^2 + 2\,(\omega/c)^2 \\
&\quad - \tfrac{4\pi G}{c^4}\,(\mu + 3p) &(18)\\
(\omega/c)^{\cdot}/c &= -\tfrac{2}{3}\,(\Theta/c)\,(\omega/c) + \tfrac{2}{\sqrt{3}}\,(\sigma/c)\,(\omega/c) + \tfrac{k}{2}\,\left(\dot{u}/c^2\right) &(19)\\
(\sigma/c)^{\cdot}/c &= \tfrac{1}{\sqrt{3}}\,\left(\dot{u}/c^2\right)' - \tfrac{a}{2\sqrt{3}}\,\left(\dot{u}/c^2\right) + \tfrac{1}{\sqrt{3}}\,\left(\dot{u}/c^2\right)^2 - \tfrac{1}{\sqrt{3}}\,(\omega/c)^2 - \tfrac{1}{\sqrt{3}}\,(\sigma/c)^2 \\
&\quad - \tfrac{2}{3}\,(\Theta/c)\,(\sigma/c) - E &(20)
\end{aligned}$$

### 2.1.2 Constraint equations

$$\begin{aligned}
0 &= (\Theta/c)' - \sqrt{3}\,(\sigma/c)' - \tfrac{3\sqrt{3}}{2}\,a\,(\sigma/c) - \tfrac{3}{2}\,k\,(\omega/c) &(21)\\
0 &= (\omega/c)' + a\,(\omega/c) - \left(\dot{u}/c^2\right)(\omega/c) &(22)\\
H &= -\sqrt{3}\,\left(\dot{u}/c^2\right)(\omega/c) + \tfrac{\sqrt{3}}{2}\,a\,(\omega/c) + \tfrac{3}{2}\,k\,(\sigma/c) &(23)
\end{aligned}$$

## 2.2 The Bianchi identities

### 2.2.1 Time derivative equations

$$\begin{aligned}
\dot{E}/c &= -\tfrac{4\pi G}{c^4}\,(\mu + p)\,(\sigma/c) + \sqrt{3}\,E\,(\sigma/c) - (\Theta/c)\,E + \tfrac{3}{2}\,k\,H &(24)\\
\dot{H}/c &= \sqrt{3}\,H\,(\sigma/c) - (\Theta/c)\,H - \tfrac{3}{2}\,k\,E &(25)\\
\dot{\mu}/c &= -\,(\mu + p)\,(\Theta/c) &(26)
\end{aligned}$$

### 2.2.2 Constraint equations

$$\begin{aligned}
E' + \tfrac{3}{2}\,a\,E &= \tfrac{4\pi G}{\sqrt{3}c^4}\,\mu' + 3H\,(\omega/c) &(27)\\
H' + \tfrac{3}{2}\,a\,H &= -\sqrt{3}\,\tfrac{4\pi G}{c^4}\,(\mu + p)\,(\omega/c) - 3E\,(\omega/c) &(28)\\
p' &= -\,(\mu + p)\,\left(\dot{u}/c^2\right) \;=\; \tfrac{\partial p}{\partial \mu}\,\mu' &(29)
\end{aligned}$$

## 2.3 Vanishing vorticity

When the vorticity vanishes, $(\omega/c) = 0$, for LRS perfect fluid spacetime geometries the Gauß equation [15], which describes the intrinsic curvature of spacelike 3-surfaces embedded orthogonal to a matter fluid flow into the spacetime manifold, can be reduced to

$$ {}^3R_{ij} = \tfrac{2}{\sqrt{3}}\,\left[\, E - \tfrac{1}{3}\,(\Theta/c)\,(\sigma/c) + \tfrac{1}{\sqrt{3}}\,(\sigma/c)^2 \,\right] e_{ij} + \tfrac{1}{3}\,\left[\, 4\,\tfrac{4\pi G}{c^4}\,\mu - \tfrac{2}{3}\,(\Theta/c)^2 + 2\,(\sigma/c)^2 \,\right] h_{ij} \ . \qquad (30)$$

The trace of this equation is

$$ {}^3R = 4\,\tfrac{4\pi G}{c^4}\,\mu - \tfrac{2}{3}\,(\Theta/c)^2 + 2\,(\sigma/c)^2 \ , \qquad (31)$$

and constitutes a generalised Friedmann equation. From this expression and the relations given by the Ricci and Bianchi identities one can also derive the conformally invariant 3-Cotton–York tensor defined by [16, 17]

$$ {}^3C^{(ij)} := -\,h^{1/3}\,\eta^{iklm}\,u_m/c\;{}^3\nabla_k\,\left[\, {}^3R^j{}_l - \tfrac{1}{4}\,h^j{}_l\,{}^3R \,\right] \ . \qquad (32)$$

## 2.4 The equations for $e^i$

Given the decomposition (15), one obtains the Ricci identity for $u^i/c$ and so the equations listed above. Analogously, given (17), one can obtain the Ricci identity for $e^i$ and a corresponding set of equations.

### 2.4.1 Ricci identities

Now from the Ricci identities we can construct

$$e^j\,\nabla_i\left(\nabla_j e^i\right) - \left(\nabla_i e^i\right)' - R_{ij}\,e^i\,e^j = 0\ , \tag{33}$$

which gives

$$0 = a' - \tfrac{2}{9}\,(\Theta/c)^2 - \tfrac{2}{3\sqrt{3}}\,(\Theta/c)\,(\sigma/c) + \tfrac{4}{3}\,(\sigma/c)^2 + \tfrac{2}{\sqrt{3}}\,E + \tfrac{a^2}{2} - \tfrac{k^2}{2} + \tfrac{4}{3}\,\tfrac{4\pi G}{c^4}\,\mu\ , \tag{34}$$

and

$$u^j/c\,\nabla_i\left(\nabla_j e^i\right) - \left(\nabla_i e^i\right)\dot{}/c - R_{ij}\,u^i/c\,e^j = 0\ , \tag{35}$$

which gives

$$0 = \dot{a}/c + \tfrac{a}{3}\,(\Theta/c) - \tfrac{a}{\sqrt{3}}\,(\sigma/c) - \tfrac{2}{3}\,(\Theta/c)\left(\dot{u}/c^2\right) + \tfrac{2}{\sqrt{3}}\,(\sigma/c)\left(\dot{u}/c^2\right) - k\,(\omega/c)\ . \tag{36}$$

### 2.4.2 Jacobi identities

Next we consider the Jacobi identities or first Bianchi identities. Here we obtain from

$$R_{[ij\,k]l}\,e^l = 0 \qquad \Leftrightarrow \qquad \nabla_{[i}\nabla_j e_{k]} = 0\ , \tag{37}$$

after contracting with $\eta^{ijkl}\,u_l/c$,

$$0 = k' + a\,k - \tfrac{2}{3}\,(\Theta/c)\,(\omega/c) - \tfrac{4}{\sqrt{3}}\,(\sigma/c)\,(\omega/c)\ , \tag{38}$$

while contracting with $\eta^{ijkl}\,e_l$, we get

$$0 = \dot{k}/c + \tfrac{1}{3}\,k\,(\Theta/c) - \tfrac{4}{\sqrt{3}}\,k\,(\sigma/c)\ . \tag{39}$$

# 3 The Consistency Equations and their Different Cases

We now turn to derive and investigate the consistency equations for the above set of covariant evolution and constraint equations for LRS perfect fluid spacetime geometries.

First we note that for any scalar invariant $f$, by the LRS symmetry,

$$\nabla_i f = f'\,e_i - \dot{f}/c\,u_i/c\ , \tag{40}$$

because the spatial derivatives in directions perpendicular to $e^i$ must be zero. Taking the covariant derivative of this relation gives

$$\nabla_j\nabla_i f = (\nabla_j f')\,e_i + f'\,(\nabla_j e_i) - \left(\nabla_j \dot{f}/c\right)\,u_i/c - \dot{f}/c\,(\nabla_j u_i/c)\ . \tag{41}$$

Multiplying by $\eta^{jikl}\,e_k\,u_l/c$, and using $(\omega/c)\,e^i = -1/2\,\eta^{ijkl}\,(\nabla_j u_k/c)\,u_l/c$ and (7), we find the important result

$$2\,\dot{f}/c\,(\omega/c) = f'\,k\ . \tag{42}$$

Applying this first to $(\omega/c)$ and using Eqs. (19) and (22), and then to $k$ and using (39) and (38), we find

$$(\omega/c)\,D = 0 = k\,D\ , \tag{43}$$

where

$$D := \tfrac{4}{3}\,(\Theta/c)\,(\omega/c) - \tfrac{4}{\sqrt{3}}\,(\sigma/c)\,(\omega/c) - a\,k\ . \tag{44}$$

Now (43) implies $D = 0$, unless $k = 0 = (\omega/c)$, but then $D$ is zero in any case, from its definition. Hence we always have

$$D = 0 \qquad \Leftrightarrow \qquad a\,k \;=\tfrac{4}{3}\,(\Theta/c)\,(\omega/c) - \tfrac{4}{\sqrt{3}}\,(\sigma/c)\,(\omega/c)\;\;. \tag{45}$$

Also applying (42) to $p$ we find

$$2\,\tfrac{\partial p}{\partial \mu}\,(\Theta/c)\,(\omega/c) = k\,\left(\dot{u}/c^2\right)\;\;. \tag{46}$$

Now to check consistency, we take the covariant time derivative along the fluid flow lines of each of the constraint equations (21), (22), (23), (27), (28) and (29) and demand that they vanish. Within this procedure we make frequent use of the relation

$$(f')^{\cdot}/c = \left(\dot{f}/c\right)' + \left(\dot{u}/c^2\right)\dot{f}/c - \tfrac{2}{\sqrt{3}}\,(\sigma/c)\,f' - \tfrac{1}{3}\,(\Theta/c)\,f'\;, \tag{47}$$

where $f$ denotes any arbitrary, covariantly defined scalar quantity, and re-substitute, where possible, for all spatial and temporal derivatives of the fluid and curvature variables from the evolution and constraint equations (18) - (29). From Eq. (27), we obtain

$$0 \;=\; \sqrt{3}\,\tfrac{4\pi G}{c^4}\,(\mu + p)\,k\;(\omega/c)\;\;, \tag{48}$$

and since we are assuming $(\mu + p) > 0$, we obtain from this

$$k\;(\omega/c) = 0\;\;. \tag{49}$$

Now we can multiply (42) first by $k$ and then by $(\omega/c)$ to attain

$$f'\,k^2 = 0 = \dot{f}/c\,(\omega/c)^2\;\;, \tag{50}$$

which is always valid. Similarly, from (46) we obtain

$$k^2\,\left(\dot{u}/c^2\right) = 0 = \tfrac{\partial p}{\partial \mu}\,(\Theta/c)\,(\omega/c)^2\;\;. \tag{51}$$

Applying (50) first to $k$ and then to $(\omega/c)$ we find that $k'\,k = 0$ and $(\omega/c)^{\cdot}/c\,(\omega/c) = 0$, which implies that always

$$k' = 0 = (\omega/c)^{\cdot}/c\;\;. \tag{52}$$

Putting the first of (51) and the second of (52) into Eq. (19) and then using (45) we find

$$a\,k = 0\;\;. \tag{53}$$

The preferred spacelike unit vector field $e^i$ in LRS spacetime geometries is thus either (i) orthogonal to spacelike 2-surfaces ($k = 0$), and can be derived from a scalar potential, or (ii) divergence-free ($a = 0$), and can be derived from a vector potential, or (iii) covariantly constant ($a = 0 = k$), when regarded in the local 3-spaces orthogonal to the matter fluid flow. Finally, putting the first of (52) in (38) and using (53), and employing the first of (51) and the second of (52) in the evolution equation (19), this shows that

$$(\Theta/c)\,(\omega/c) = 0 = (\sigma/c)\,(\omega/c)\;\;. \tag{54}$$

Using these results in the rest of the integrability conditions referred to above, the only non-trivial one is the propagation equation for the magnitude of the acceleration of the fluid,

$$0 \;=\; \left(\dot{u}/c^2\right)^{\cdot}/c - \left[\tfrac{\partial p}{\partial \mu}\,(\Theta/c)\right]' - \left(\tfrac{\partial p}{\partial \mu} - \tfrac{1}{3}\right)(\Theta/c)\left(\dot{u}/c^2\right) + \tfrac{2}{\sqrt{3}}\,(\sigma/c)\left(\dot{u}/c^2\right)\;\;, \tag{55}$$

which implies no new constraints (in checking these conditions, note that (54) implies $(\omega/c)\left[\tfrac{\partial p}{\partial \mu}\,(\Theta/c)\right]' = 0$).

Hence, the set of covariant evolution and constraint equations describing LRS perfect fluid spacetime geometries with $(\mu + p) > 0$ is consistent provided that the conditions

$$k\,(\omega/c) = 0\;, \qquad a\,k = 0 = (\Theta/c)\,(\omega/c) = (\sigma/c)\,(\omega/c)\;\;, \qquad \dot{f}/c\,(\omega/c) = 0 = f'\,k\;\;, \tag{56}$$

are true simultaneously. (The case $\dot{f}/c = \left(\dot{u}/c^2\right)$ is *not* included in this notation, as $\left(\dot{u}/c^2\right)$ is *not* the time derivative of a covariantly defined scalar.) Then also the covariant time derivatives of the constraint equations (34) and (38) vanish identically. Assuming $(\mu + p) > 0$, there are thus three cases that can occur [5, 6]:

- **LRS class I**: $(\omega/c) \neq 0 \quad \Rightarrow k = (\Theta/c) = (\sigma/c) = 0$ , $\dot{f}/c = 0$ .
  $e^i$ is hypersurface orthogonal, $u^i/c$ is twisting.
- **LRS class II**: $k = 0 = (\omega/c)$ .
  $e^i$ and $u^i/c$ *both* are hypersurface orthogonal.
- **LRS class III**: $k \neq 0 \quad \Rightarrow (\omega/c) = a = (\dot{u}/c^2) = 0$ , $f' = 0$ .
  $e^i$ is twisting, $u^i/c$ is hypersurface orthogonal.

We discuss these in turn, in each case considering first the generic situation and then the subcases that occur.

It remains to check the consistency of the algebraic expression (23) for $H$ with the constraint equation (28). When this is done and the conditions (56) have been imposed, one obtains

$$0 = \Big[\, \sqrt{3}\,(\dot{u}/c^2)' + \sqrt{3}\,(\dot{u}/c^2)^2 - \tfrac{4\pi G}{\sqrt{3}c^4}\,(\mu + 3p) - 2\,E\,\Big]\,(\omega/c) \ . \tag{57}$$

This is only of interest when $(\omega/c) \neq 0$ (see the following section).

# 4 $(\omega/c) \neq 0$: Rotating Solutions (LRS Class I)

When $(\omega/c) \neq 0$, we immediately find that $\Rightarrow k = 0 = (\Theta/c) = (\sigma/c)$, so the models within this LRS class of solutions can neither expand nor distort. Further, the consistency conditions (56) show that $\dot{f}/c = 0$ for *all* covariantly defined scalars $f$, that is $\partial_{ct}$ is a timelike Killing vector field and all spacetimes within this LRS class will also be stationary. Thus, in this case there exists a $G_4$ multiply-transitive on timelike 3-surfaces. Non-zero quantities in general are $\mu$, $p$, $(\dot{u}/c^2)$, $a$, $E$ and $H$.

The set of equations one needs to solve in this LRS class is given by

$$\begin{aligned}
(\dot{u}/c^2)' &= -\,a\,(\dot{u}/c^2) - (\dot{u}/c^2)^2 - 2\,(\omega/c)^2 + \tfrac{4\pi G}{c^4}\,(\mu + 3p) && (58)\\
(\omega/c)' &= -\,a\,(\omega/c) + (\dot{u}/c^2)\,(\omega/c) && (59)\\
(\dot{u}/c^2) &= -\,\tfrac{\partial p}{\partial \mu}\,\mu'\,/\,(\mu + p) && (60)\\
a' &= -\,\tfrac{a^2}{2} + a\,(\dot{u}/c^2) + 2\,(\omega/c)^2 - 2\,\tfrac{4\pi G}{c^4}\,(\mu + p) \ , && (61)
\end{aligned}$$

which follow from (18), (22), (29) and (34) respectively. The magnitudes of the "electric" and "magnetic parts" of the Weyl conformal curvature tensor in these models are given algebraically by a combination of Eqs. (20) and (18) and Eq. (23) respectively, which are

$$\begin{aligned}
E &= -\,\tfrac{\sqrt{3}}{2}\,a\,(\dot{u}/c^2) - \sqrt{3}\,(\omega/c)^2 + \tfrac{4\pi G}{\sqrt{3}c^4}\,(\mu + 3p) && (62)\\
H &= -\,\sqrt{3}\,(\dot{u}/c^2)\,(\omega/c) + \tfrac{\sqrt{3}}{2}\,a\,(\omega/c) \ . && (63)
\end{aligned}$$

With Eqs. (58) and (62) Eq. (57) is identically satisfied, as are the constraint equations (27) and (28) on substitution of (62) and (63).

When the acceleration is non-zero, by combining Eq. (58) with (60) we can derive a coupled second-order ordinary differential equation for the spatial distribution of the energy density, which is

$$\begin{aligned}
&\tfrac{\partial p}{\partial \mu}\,\mu'' + \tfrac{\partial^2 p}{\partial \mu^2}\,\mu'^2 + a\,\tfrac{\partial p}{\partial \mu}\,\mu' - \Big(2\,\tfrac{\partial p}{\partial \mu} + 1\Big)\,\tfrac{\partial p}{\partial \mu}\,\mu'^2/\,(\mu + p)\\
&\qquad - 2\,(\mu + p)\,(\omega/c)^2 + \tfrac{4\pi G}{c^4}\,(\mu + p)\,(\mu + 3p) \quad = \quad 0 \ . && (64)
\end{aligned}$$

**Algorithm:** The *boundary data*, which can be specified *freely* on a timelike 3-surface orthogonal to $e^i$, are the values of $\mu$, $(\omega/c)$ and $a$. We assume a given equation of state $p(\mu)$. $(\dot{u}/c^2)$ is then given by Eq. (60), while $E$ and $H$ follow from Eqs. (62) and (63).

For a treatment of this LRS class in terms of local coordinates and a metric tensor field refer to Stewart and Ellis [6], Eq. (2.8), and Kramer et al [1], Eq. (11.4).

### 4.1 Solutions with $a=0$

A subclass of solutions with $a=0$ exists. However, for consistency they demand an equation of state of the form $p(\mu)=-1/3\,\mu+\text{const}$, which is usually dismissed as unphysical. One ordinary differential equation describing the spatial distribution of the total energy density $\mu$ remains to be solved. It follows from Eq. (58) and is given by

$$\mu''-\tfrac{1}{3}\,\mu'^{2}\,/\,(\mu+p)+3\,\tfrac{4\pi G}{c^4}\,(\mu^2-p^2)=0\ . \tag{65}$$

One then obtains the following expressions for the remaining non-zero quantities:

$$\begin{array}{rclrcl}(\dot u/c^2) &=& \tfrac{1}{3}\,\mu'\,/\,(\mu+p) & (\omega/c) &=& \left(\tfrac{4\pi G}{c^4}\right)^{1/2}(\mu+p)^{1/2}\\ E &=& -\tfrac{2}{\sqrt{3}}\,\tfrac{4\pi G}{c^4}\,\mu & H &=& -\left(\tfrac{4\pi G}{3c^4}\right)^{1/2}\,\mu'\,/\,(\mu+p)^{1/2}\ .\end{array} \tag{66}$$

Models of this kind can be regarded as (non-physical) generalisations of the Gödel LRS case (see subsection 4.5 below).

### 4.2 Solutions with $p=0$

In the subclass of dust models, $p=0\Leftrightarrow(\dot u/c^2)=0$, deriving from Eq. (61) there is again one ordinary differential equation for $\mu$ which remains to be solved:

$$\mu''-\tfrac{5}{4}\,\mu'^{2}\,/\mu-2\,\tfrac{4\pi G}{c^4}\,\mu^2=0\ . \tag{67}$$

Then it follows that

$$\begin{array}{rclrcl}a &=& -\tfrac{1}{2}\,\mu'\,/\mu & (\omega/c) &=& \left(\tfrac{2\pi G}{c^4}\right)^{1/2}\,\mu^{1/2}\\ E &=& -\tfrac{2\pi G}{\sqrt{3}c^4}\,\mu & H &=& -\tfrac{\sqrt{3}}{4}\left(\tfrac{2\pi G}{c^4}\right)^{1/2}\,\mu'\,/\mu^{1/2}\ .\end{array} \tag{68}$$

Models with these properties can also be regarded as generalisations of the Gödel LRS case (see subsection 4.5 below).

If we were to impose the dynamical restriction $(\dot u/c^2)=0$ instead of $p=0$, Eq. (60) would be solved by either $\mu'=0$, which leads to $f'=0$ (subsection 4.5 below), or $\frac{\partial p}{\partial\mu}=0\Rightarrow p=\text{const}$, which leads to a slight generalisation of Eq. (67).

### 4.3 Solutions with $H=0$

The dynamical restriction $H=0$ implies from Eq. (63) that $a=2\,(\dot u/c^2)$. Then from Eqs. (58) and (61) we find that

$$(\omega/c)^2=-\,(\dot u/c^2)^2+\tfrac{2}{3}\,\tfrac{4\pi G}{c^4}\,(\mu+2p)\ . \tag{69}$$

Finally, consistency of Eq. (59), on using Eq. (60), demands that $\frac{\partial p}{\partial\mu}=1\Rightarrow p(\mu)=\mu+\text{const}$.

### 4.4 Solutions with $E=0$

The dynamical restriction $E=0$ implies from Eq. (62) that

$$(\omega/c)^2=-\tfrac{a}{2}\,(\dot u/c^2)+\tfrac{4\pi G}{3c^4}\,(\mu+3p)\ . \tag{70}$$

Consistency with Eq. (59) then requires that

$$(\dot u/c^2)^2-\left[\,\tfrac{a}{2}+\tfrac{4}{3\,a}\,\tfrac{4\pi G}{c^4}\,p+\tfrac{2}{3\,a}\,[\,\tfrac{1}{3}\left(\tfrac{\partial p}{\partial\mu}\right)^{-1}+1\,]\,\tfrac{4\pi G}{c^4}\,(\mu+p)\,\right](\dot u/c^2)+\tfrac{4\pi G}{3c^4}\,(\mu+3p)\ =\ 0\ , \tag{71}$$

in order to obtain a purely "magnetic" solution.

### 4.5 $f' = 0$: Gödel's rotating model of the Universe

If we impose the additional condition that $f' = 0$, the symmetry group is a $G_5$ multiply-transitive on the full spacetime manifold, which consequently is homogeneous. As the spatial derivatives of *all* non-zero scalar quantities $f$ vanish, we have immediately from Eq. (60) that the matter moves geodesically: $(\dot{u}/c^2) = 0$. However, since *no* spacelike 3-surfaces of constant $f$ exist, which are the group orbits of a simply-transitive $G_3$, models of this LRS subclass are *not* spatially homogeneous (see Section 5 below). Equation (59) shows that $a = 0$, and from the algebraic equation (63) we obtain that the magnitude of the "magnetic part" of the Weyl conformal curvature tensor is zero, $H = 0$. All scalar quantities $f$ are constant on the spacetime manifold, so the remaining equations are purely algebraic. From Eqs. (58), (61) and (62) we find

$$
\begin{aligned}
2\,(\omega/c)^2 &= \tfrac{4\pi G}{c^4}\,(\mu + 3p) \;=\; 2\,\tfrac{4\pi G}{c^4}\,(\mu + p) && (72)\\
E &= -\sqrt{3}\,(\omega/c)^2 + \tfrac{4\pi G}{\sqrt{3}c^4}\,(\mu + 3p)\ . && (73)
\end{aligned}
$$

It can easily be seen that this system of algebraic equations is consistent only, provided the equation of state is $p(\mu) = \mu$, that is for stiff matter. An equivalent configuration would be a combination of dust matter, $p = 0$, and a *negative* cosmological constant, $\Lambda < 0$ [15], which gives Gödel's rotating model of the Universe [18].

In summary, we observe that LRS class I of perfect fluid spacetime geometries with equation of state $p = p(\mu)$ contains a variety of stationary solutions, that, apart from the Gödel model, are differentially rotating. Most of them, however, are of minor interest for astrophysical and cosmological purposes, as it is difficult to see how, given non-zero vorticity, the geometry of a star model could correspond to exact rotational symmetry about every point.

## 5 $k \neq 0$: Homogeneous Orthogonal Models with Twist (LRS Class III)

When $k \neq 0$, the consistency conditions (56) demand that $f' = 0$, $(\omega/c) = 0 = a$: *all* spatial derivatives vanish and it follows immediately from Eq. (29) that the matter in these models moves on timelike geodesics, that is $(\dot{u}/c^2) = 0$. Thus, since $u^i/c$ is normal and geodesic, all scalars $f$ are spatially homogeneous and there exists a $G_4$ of isometries multiply-transitive on spacelike 3-surfaces orthogonal to $u^i/c$. That is, the spacetimes themselves are (orthogonally) spatially homogeneous (OSH) [7]. The non-zero quantities in the generic case are $\mu$, $p$, $(\Theta/c)$, $(\sigma/c)$, $E$ and $H$. Note that, apart from the class-defining spatial rotation of $e^i$, all vectorial quantities are zero. From the 3-Ricci identities and the Gauß equation for $e^i$ [15] the 3-Ricci curvature tensor of the spacelike 3-surfaces orthogonal to $u^i/c$ can be determined to be

$$
\begin{aligned}
{}^3R_{ij} &= \tfrac{2}{\sqrt{3}}\left[\,E - \tfrac{1}{3}\,(\Theta/c)\,(\sigma/c) + \tfrac{1}{\sqrt{3}}\,(\sigma/c)^2\,\right] e_{ij}\\
&\quad + \tfrac{1}{3}\left[\,\tfrac{3}{2}\,k^2 - 2\sqrt{3}\,E + \tfrac{2}{\sqrt{3}}\,(\Theta/c)\,(\sigma/c) - 2\,(\sigma/c)^2\,\right] h_{ij}\ , && (74)
\end{aligned}
$$

the trace of which, when combined with Eq. (31), yields the generalised Friedmann equation

$$
\begin{aligned}
{}^3R &= \tfrac{3}{2}\,k^2 - 2\sqrt{3}\,E + \tfrac{2}{\sqrt{3}}\,(\Theta/c)\,(\sigma/c) - 2\,(\sigma/c)^2\\
&= 4\,\tfrac{4\pi G}{c^4}\,\mu - \tfrac{2}{3}\,(\Theta/c)^2 + 2\,(\sigma/c)^2\ . && (75)
\end{aligned}
$$

The magnitudes of the "electric" and "magnetic parts" of the Weyl conformal curvature tensor are determined algebraically from Eqs. (75) or (34) and Eq. (23) respectively as

$$
\begin{aligned}
E &= \tfrac{1}{3\sqrt{3}}\,(\Theta/c)^2 + \tfrac{1}{3}\,(\Theta/c)\,(\sigma/c) - \tfrac{2}{\sqrt{3}}\,(\sigma/c)^2 + \tfrac{\sqrt{3}}{4}\,k^2 - 2\,\tfrac{4\pi G}{\sqrt{3}c^4}\,\mu && (76)\\
H &= \tfrac{3}{2}\,k\,(\sigma/c)\ , && (77)
\end{aligned}
$$

and a time evolution equation for the dynamical variable $k$ follows from Eq. (39). We emphasise the fact that throughout this LRS class $H \neq 0$, unless we deal with a spatially isotropic situation (see subsection 5.3 below). Using Eqs. (25) and (77) one can show that the conformal 3-Cotton–York tensor in this LRS class is given by (see also [19])

$$ {}^3C_{ij} = -\tfrac{2}{\sqrt{3}}\,h^{1/3}\,\Big[\,\dot{H}/c + \tfrac{4}{3}\,H\,(\Theta/c) - \tfrac{4}{\sqrt{3}}\,H\,(\sigma/c)\,\Big]\,e_{ij}\ . \tag{78} $$

The set of dynamical equations describing this class of expanding LRS models reads

$$ \begin{aligned} (\Theta/c)\dot{}/c &= -\tfrac{1}{3}\,(\Theta/c)^2 - 2\,(\sigma/c)^2 - \tfrac{4\pi G}{c^4}\,(\mu+3p) && (79)\\ (\sigma/c)\dot{}/c &= -\tfrac{1}{3\sqrt{3}}\,(\Theta/c)^2 - (\Theta/c)\,(\sigma/c) + \tfrac{1}{\sqrt{3}}\,(\sigma/c)^2 - \tfrac{\sqrt{3}}{4}\,k^2 + 2\,\tfrac{4\pi G}{\sqrt{3}c^4}\,\mu && (80)\\ \dot{\mu}/c &= -\,(\mu+p)\,(\Theta/c) && (81)\\ \dot{k}/c &= -\tfrac{1}{3}\,k\,(\Theta/c) + \tfrac{4}{\sqrt{3}}\,k\,(\sigma/c)\ . && (82) \end{aligned} $$

We have tested for consistency of this set of equations with the evolution equation (24), which, with Eq. (76), is identically satisfied. An evolution equation for the 3-Ricci scalar, Eq. (75), can be derived:

$$ [\,{}^3R\,]\dot{}/c = -\tfrac{2}{3}\,(\Theta/c)\,{}^3R + \tfrac{2}{\sqrt{3}}\,(\sigma/c)\,{}^3R - \sqrt{3}\,k^2\,(\sigma/c)\ . \tag{83} $$

*Alternatively*, in the face of our aim of using the set $S_4 = \{\,\mu, (\Theta/c)\,, (\sigma/c)\,, E\,\}$ as the (maximum set of) four generalised essential coordinates for the investigation of functional dependencies in models with non-zero rate of expansion, one can solve Eq. (34) for $k^2$ instead, obtaining

$$ k^2 = -\tfrac{4}{9}\,(\Theta/c)^2 - \tfrac{4}{3\sqrt{3}}\,(\Theta/c)\,(\sigma/c) + \tfrac{8}{3}\,(\sigma/c)^2 + \tfrac{4}{\sqrt{3}}\,E + \tfrac{8}{3}\,\tfrac{4\pi G}{c^4}\,\mu\ . \tag{84} $$

As $k^2 > 0$ is demanded, this leads to an algebraic restriction on the values of $\mu$, $(\Theta/c)$, $(\sigma/c)$ and $E$. The set of dynamical equations is then

$$ \begin{aligned} (\Theta/c)\dot{}/c &= -\tfrac{1}{3}\,(\Theta/c)^2 - 2\,(\sigma/c)^2 - \tfrac{4\pi G}{c^4}\,(\mu+3p) && (85)\\ (\sigma/c)\dot{}/c &= -\tfrac{1}{\sqrt{3}}\,(\sigma/c)^2 - \tfrac{2}{3}\,(\Theta/c)\,(\sigma/c) - E && (86)\\ \dot{E}/c &= \tfrac{4\pi G}{c^4}\,(5\mu - p)\,(\sigma/c) + 4\sqrt{3}\,E\,(\sigma/c) - (\Theta/c)\,E \\ &\quad -\,(\Theta/c)^2\,(\sigma/c) - \sqrt{3}\,(\Theta/c)\,(\sigma/c)^2 + 6\,(\sigma/c)^3 && (87)\\ \dot{\mu}/c &= -\,(\mu+p)\,(\Theta/c)\ . && (88) \end{aligned} $$

We have tested for consistency of this set of equations with the evolution equation (82), which, with Eq. (84), is identically satisfied.

**Algorithm:** The *initial data*, which can be specified *freely* on a spacelike 3-surface orthogonal to the matter flow, are the values of $\mu$, $(\Theta/c)$, $(\sigma/c)$ and $E$. Given the equation of state $p(\mu)$, all time derivatives are determined as well as $k$ from Eq. (84) and subsequently $H$ from Eq. (77).

If local comoving coordinates are chosen, the line element can be cast into the following form (cf. Eq. (2.8) of Stewart and Ellis [6] and Eq. (11.4) of Kramer et al [1]):

$$ ds^2 = X^2(ct)\,[\,dx - h(y)\,dz\,]^2 + Y^2(ct)\,[\,dy^2 + \Sigma^2(y)\,dz^2\,] - d(ct)^2\ . \tag{89} $$

Then one immediately obtains relations for the following scalars:

$$ (\Theta/c) = \frac{\dot{X}/c}{X} + 2\,\frac{\dot{Y}/c}{Y} \qquad (\sigma/c) = \frac{1}{\sqrt{3}}\left(\frac{\dot{X}/c}{X} - \frac{\dot{Y}/c}{Y}\right)\ . \tag{90} $$

This enables one to integrate Eq. (82), which gives

$$ k = C_1\,\frac{X}{Y^2}\ , \tag{91} $$

where $C_1$ denotes an integration constant. The solution to Eq. (83) is given by

$$ {}^3R = 2\,\frac{C_2}{Y^2} - \frac{{C_1}^2}{2}\,\frac{X^2}{Y^4}\ , \tag{92} $$

where $C_2$ denotes a further integration constant, and thus Eq. (75) constitutes a first integral. Consequently, one is left with only three essential dynamical equations, these being Eqs. (75), (80) and (81). Equivalent is the set of equations

$$2\,\frac{\ddot{Y}/c^2}{Y} + \left(\frac{\dot{Y}/c}{Y}\right)^2 + \frac{C_2}{Y^2} - \frac{3\,{C_1}^2}{4}\,\frac{X^2}{Y^4} \;=\; -\,2\,\tfrac{4\pi G}{c^4}\,p \tag{93}$$

$$2\,\frac{\dot{X}/c}{X}\,\frac{\dot{Y}/c}{Y} + \left(\frac{\dot{Y}/c}{Y}\right)^2 + \frac{C_2}{Y^2} - \frac{{C_1}^2}{4}\,\frac{X^2}{Y^4} \;=\; 2\,\tfrac{4\pi G}{c^4}\,\mu \tag{94}$$

$$\frac{\ddot{X}/c^2}{X} + \frac{\dot{X}/c}{X}\,\frac{\dot{Y}/c}{Y} + \frac{\ddot{Y}/c^2}{Y} + \frac{{C_1}^2}{4}\,\frac{X^2}{Y^4} \;=\; -\,2\,\tfrac{4\pi G}{c^4}\,p\;. \tag{95}$$

This set of equations was given in Ref. [5] for $p = 0$. Additionally we want to point out that, in contrast to $C_1$, *no* invariant geometric meaning can be associated with the constant $C_2$. In fact $C_2$ is just a constant of integration, resulting from the fact that (94) is a first integral of (93) and (95); so any two of these equations implies the third. One obtains a form of the field equations independent of $C_2$ from (95) and the difference between (93) and (94).

Exact solutions to the dynamical equations within this LRS class have been discussed in Kramer et al, chapter 12.3 [1]. The different underlying simply-transitive subgroups $G_3$ of the possible $G_4$ isometry groups can be of Bianchi Type–II, Type–VIII/III and Type–IX [7]. For Type–IX one *can* have ${}^3R > 0$, which means ${}^3R$ *can* change its sign in this class of models. Type–II and Type–VIII/III have ${}^3R < 0$ throughout their entire evolution and are only distinguished by different values of the sectional 3-curvature. Dynamically they are identical.

## 5.1 Dust solutions

Because the fluid flow is geodesic in any case, the dust subcases are simply distinguished by marginally simpler evolution equations ($p = 0$ in Eqs. (81), (93) and (95)).

## 5.2 Solutions with $E = 0$ ("Pure magnetic")

Imposing the dynamical restriction $E = 0$, Eq. (87) yields for $(\sigma/c) \neq 0$ the algebraic condition

$$(\Theta/c)^2 = \tfrac{4\pi G}{c^4}\,(5\mu - p) - \sqrt{3}\,(\Theta/c)\,(\sigma/c) + 6\,(\sigma/c)^2\;\;, \tag{96}$$

which in order to satisfy Eq. (85) restricts the equation of state through an algebraic condition for $\frac{\partial p}{\partial \mu}$. The value of $k^2$, as determined by Eq. (84), is

$$k^2 = \tfrac{4}{9}\,\tfrac{4\pi G}{c^4}\,(\mu + p)\;\;, \tag{97}$$

which is clearly consistent with the condition $k^2 > 0$, and thus with $(\sigma/c) \neq 0$ we have from Eq. (77) that $H \neq 0$. Solutions of this kind are of purely "magnetic" character as regards the Weyl conformal curvature tensor. An example is provided by the self-similar Bianchi Type–II OSH LRS solutions of Collins and Stewart [20], which have a linear barotropic equation of state of the form $p(\mu) = (\gamma - 1)\mu$, and one obtains $E = 0$ for $\gamma = 6/5$ (see also [21]).

## 5.3 The FLRW subcase

If we demand that for $(\sigma/c) \neq 0$ the spacelike 3-surfaces orthogonal to $u^i/c$ be of *constant* curvature, that is from Eq. (74) that

$$E = \tfrac{1}{3}\,[\;(\Theta/c) - \sqrt{3}\,(\sigma/c)\;]\;(\sigma/c)\;\;, \tag{98}$$

one obtains from Eq. (87) an algebraic expression for $(\Theta/c)^2$, which, when substituted in Eq. (84), gives $k^2 = 0$ and thus violates the definition of this specific LRS class. Thus there do *not* exist any shearing solutions in this LRS class with isotropic ${}^3R_{ij}$; the ${}^3R > 0$ - FLRW models are the only LRS models with $k \neq 0$ and 3-spaces of constant curvature, as follows from Eq. (75). These are invariant under a $G_6$ of isometries multiply-transitive on spacelike 3-surfaces. In this case there exists a 3-D family of rotational symmetries rather than one. This is the exceptional case where the spacelike unit vector field $e^i$ is *not* uniquely defined, and there is *no* covariant feature that picks out a preferred spatial direction. Nevertheless, we can still find local coordinate or orthonormal frame bases as before. In the local comoving coordinates of Eq. (89) we have that $Y = X$.

The Einstein static model is the special FLRW case when $(\Theta/c) = 0$, and thus $k^2 = 8/3\,\frac{4\pi G}{c^4}\,\mu > 0$ from Eq. (84). Then the spacetime manifold is invariant under a multiply-transitive $G_7$ of isometries and therefore homogeneous. From Eq. (85) we find that for consistency an equation of state of the form $p(\mu) = -1/3\,\mu = \text{const}$ is required, which can be interpreted as a dust model ($p = 0$) with a *positive* cosmological constant, $\Lambda > 0$.

In summary, the general OSH LRS class III perfect fluid spacetimes with equation of state $p = p(\mu)$ are characterised by the existence of a $G_4$ isometry group multiply-transitive on spacelike 3-surfaces, with a simply-transitive subgroup $G_3$ which belongs to one of the various Bianchi types. Special cases are the ${}^3R > 0$ - FLRW models. Of particular interest is the possibility of constructing simple cosmological models with purely "magnetic" Weyl conformal curvature tensor. This offers a step to studying the underlying physical mechanisms which could generate solutions of this peculiar kind.

# 6 $k = 0 = (\omega/c)$: The Inhomogeneous Orthogonal Family (LRS Class II)

When $k = 0 = (\omega/c)$, there exist 3-surfaces orthogonal to the fluid flow in which there acts a $G_3$ multiply-transitive on spacelike 2-surfaces orthogonal to $e^i$. These are the spherically symmetric solutions and their generalisations with plane and hyperbolic 2-spaces. All members within this class of expanding (or in the time-reversed case contracting), spatially inhomogeneous LRS models have vanishing "magnetic part" of the Weyl conformal curvature tensor, as follows directly from the constraint equation (23), that is

$$H = 0\ . \tag{99}$$

The non-zero quantities in the generic case are $\mu$, $p$, $(\Theta/c)$, $(\sigma/c)$, $(\dot{u}/c^2)$, $a$, and $E$. Depending on which of these dynamical variables might be zero, a broad variety of different special cases arises, which we will discuss in detail in this and the following section.

From the 3-Ricci identities and the Gauß equation for $e^i$ [15] the 3-Ricci curvature tensor of the spacelike 3-surfaces orthogonal to $u^i/c$ can be determined to be

$${}^3R_{ij} = -\tfrac{1}{3}\,[\ a' + 2\,K\ ]\ e_{ij} - \tfrac{1}{3}\left[\ 2\,a' + \tfrac{3}{2}\,a^2 - 2\,K\ \right]\ h_{ij}\ , \tag{100}$$

where $K$ denotes the (constant) Gaußian curvature ${}^2R := 2\,K$ of the 2-D spacelike group orbits orthogonal to $e^i$ and $u^i/c$. A further constraint is given by the generalised Friedmann equation

$$\begin{aligned} {}^3R &= -\left[\ 2\,a' + \tfrac{3}{2}\,a^2 - 2\,K\ \right] \\ &= 4\,\tfrac{4\pi G}{c^4}\,\mu - \tfrac{2}{3}\,(\Theta/c)^2 + 2\,(\sigma/c)^2\ . \end{aligned} \tag{101}$$

From this we obtain for $a'$ the expression

$$a' = -\tfrac{3}{4}\,a^2 + K - 2\,\tfrac{4\pi G}{c^4}\,\mu + \tfrac{1}{3}\,(\Theta/c)^2 - (\sigma/c)^2\ . \tag{102}$$

This relation can be combined with Eq. (34) to give a purely algebraic expression for $E$, which is

$$E = \tfrac{4\pi G}{\sqrt{3}c^4}\,\mu + \tfrac{\sqrt{3}}{8}\,a^2 - \tfrac{\sqrt{3}}{2}\,K - \tfrac{1}{6\sqrt{3}}\,[\ (\Theta/c) - \sqrt{3}\,(\sigma/c)\ ]^2\ . \tag{103}$$

Then the set of equations describing all LRS models within this class, the metric tensor of which can always be diagonalised [1], is

$$\begin{aligned} (\Theta/c)^{\cdot}/c &= -\tfrac{1}{3}\,(\Theta/c)^2 + \left(\dot{u}/c^2\right)' + a\left(\dot{u}/c^2\right) + \left(\dot{u}/c^2\right)^2 - 2\,(\sigma/c)^2 - \tfrac{4\pi G}{c^4}\,(\mu + 3p) \qquad (104) \\ (\sigma/c)^{\cdot}/c &= \tfrac{1}{\sqrt{3}}\left(\dot{u}/c^2\right)' - \tfrac{a}{2\sqrt{3}}\left(\dot{u}/c^2\right) + \tfrac{1}{\sqrt{3}}\left(\dot{u}/c^2\right)^2 + \tfrac{1}{6\sqrt{3}}\,(\Theta/c)^2 \end{aligned}$$

$$
\begin{aligned}
& - (\Theta/c)\,(\sigma/c) - \tfrac{1}{2\sqrt{3}}\,(\sigma/c)^2 - \tfrac{4\pi G}{\sqrt{3}c^4}\,\mu - \tfrac{\sqrt{3}}{8}\,a^2 + \tfrac{\sqrt{3}}{2}\,K && (105)\\
\dot{\mu}/c &= -\,(\mu+p)\,(\Theta/c) && (106)\\
\dot{a}/c &= -\tfrac{a}{3}\,(\Theta/c) + \tfrac{a}{\sqrt{3}}\,(\sigma/c) + \tfrac{2}{3}\,(\Theta/c)\,(\dot{u}/c^2) - \tfrac{2}{\sqrt{3}}\,(\sigma/c)\,(\dot{u}/c^2) && (107)\\
\dot{K}/c &= -\tfrac{2}{3}\,K\,(\Theta/c) + \tfrac{2}{\sqrt{3}}\,K\,(\sigma/c) && (108)\\
(\sigma/c)' &= \tfrac{1}{\sqrt{3}}\,(\Theta/c)' - \tfrac{3}{2}\,a\,(\sigma/c) && (109)\\
(\dot{u}/c^2) &= -\tfrac{\partial p}{\partial \mu}\,\mu'\,/\,(\mu+p) && (110)\\
a' &= -\tfrac{3}{4}\,a^2 + K - 2\,\tfrac{4\pi G}{c^4}\,\mu + \tfrac{1}{3}\,(\Theta/c)^2 - (\sigma/c)^2 && (111)\\
K' &= -\,a\,K && (112)
\end{aligned}
$$

We have tested for consistency of this set of equations with the evolution equation (24), which, on using Eq. (103), is identically satisfied. The evolution equation (108) derives from demanding preservation in time along the matter flow lines of the constraint (102), whereas expression (112) is a consequence of the constraint (27). The covariant time derivative of the constraint (112) vanishes identically.

By application of Eq. (112) one can show that the conformal 3-Cotton–York tensor for all models within this LRS class vanishes (see also [19]),

$$ {}^3C_{ij} = 0\ , \qquad (113) $$

that is, the spacelike 3-surfaces orthogonal to $u^i/c$ are conformally flat. The evolution equation for the 3-Ricci scalar (101) is

$$
\begin{aligned}
[\,{}^3R\,]^{\cdot}/c &= -\tfrac{1}{3}\,[\,2\,(\Theta/c) + \sqrt{3}\,(\sigma/c)\,]\,[\,{}^3R + 2\,a\,(\dot{u}/c^2)\,] + 2\sqrt{3}\,[\,K - \tfrac{a^2}{4}\,]\,(\sigma/c)\\
&\quad - \tfrac{4}{3}\,[\,(\Theta/c) - \sqrt{3}\,(\sigma/c)\,]\,[\,(\dot{u}/c^2)' + (\dot{u}/c^2)^2\,]\ . \qquad (114)
\end{aligned}
$$

*Alternatively*, in the face of our aim of using the set $S_4 = \{\,\mu, (\Theta/c)\,, (\sigma/c)\,, E\,\}$ as the (maximum set of) four generalised essential coordinates for the investigation of functional dependencies in models with non-zero rate of expansion, one can solve the combined Eqs. (34) and (102) for $K$ instead, obtaining

$$ K = \tfrac{2}{3}\,\tfrac{4\pi G}{c^4}\,\mu + \tfrac{a^2}{4} - \tfrac{2}{\sqrt{3}}\,E - \tfrac{1}{9}\,[\,(\Theta/c) - \sqrt{3}\,(\sigma/c)\,]^2\ . \qquad (115) $$

Then the set of relevant dynamical equations is

$$
\begin{aligned}
(\Theta/c)^{\cdot}/c &= -\tfrac{1}{3}\,(\Theta/c)^2 + (\dot{u}/c^2)' + a\,(\dot{u}/c^2) + (\dot{u}/c^2)^2 - 2\,(\sigma/c)^2 - \tfrac{4\pi G}{c^4}\,(\mu+3p) && (116)\\
(\sigma/c)^{\cdot}/c &= \tfrac{1}{\sqrt{3}}\,(\dot{u}/c^2)' - \tfrac{a}{2\sqrt{3}}\,(\dot{u}/c^2) + \tfrac{1}{\sqrt{3}}\,(\dot{u}/c^2)^2 - \tfrac{1}{\sqrt{3}}\,(\sigma/c)^2 - \tfrac{2}{3}\,(\Theta/c)\,(\sigma/c) - E && (117)\\
\dot{E}/c &= -\tfrac{4\pi G}{c^4}\,(\mu+p)\,(\sigma/c) + \sqrt{3}\,E\,(\sigma/c) - (\Theta/c)\,E && (118)\\
\dot{\mu}/c &= -\,(\mu+p)\,(\Theta/c) && (119)\\
\dot{a}/c &= -\tfrac{a}{3}\,(\Theta/c) + \tfrac{a}{\sqrt{3}}\,(\sigma/c) + \tfrac{2}{3}\,(\Theta/c)\,(\dot{u}/c^2) - \tfrac{2}{\sqrt{3}}\,(\sigma/c)\,(\dot{u}/c^2) && (120)\\
(\sigma/c)' &= \tfrac{1}{\sqrt{3}}\,(\Theta/c)' - \tfrac{3}{2}\,a\,(\sigma/c) && (121)\\
E' &= -\tfrac{3}{2}\,a\,E + \tfrac{4\pi G}{\sqrt{3}c^4}\,\mu' && (122)\\
(\dot{u}/c^2) &= -\tfrac{\partial p}{\partial \mu}\,\mu'\,/\,(\mu+p) && (123)\\
a' &= \tfrac{2}{9}\,(\Theta/c)^2 + \tfrac{2}{3\sqrt{3}}\,(\Theta/c)\,(\sigma/c) - \tfrac{4}{3}\,(\sigma/c)^2 - \tfrac{2}{\sqrt{3}}\,E - \tfrac{a^2}{2} - \tfrac{4}{3}\,\tfrac{4\pi G}{c^4}\,\mu\ . && (124)
\end{aligned}
$$

We have tested for consistency of this set of equations with the evolution equation (108) and the constraint equation (112), which, with Eq. (115), are identically satisfied.

**Algorithm:** The *initial data*, which can be specified *freely* on a spacelike 3-surface orthogonal to the matter flow, are the values of $\mu$ and $(\Theta/c)$, while $(\sigma/c)$, $E$ and $a$ can be given at a point and their spatial distribution is determined by Eqs. (121), (122) and (124) respectively. Then, given a choice of the equation of state $p(\mu)$, $(\dot{u}/c^2)$ follows from Eq. (123), all time derivatives are known and $K$ is given by Eq. (115).

If local comoving coordinates are chosen, the line element can be cast into the following form (cf. Eq. (2.8) of Stewart and Ellis [6] and Eq. (13.2) of Kramer et al [1]):

$$ds^2 = X^2(x,ct)\,dx^2 + Y^2(x,ct)\,\left[\,dy^2 + \Sigma^2(y)\,dz^2\,\right] - F^{-2}(x,ct)\,d(ct)^2 \tag{125}$$

Then one immediately obtains relations for the following scalars:

$$a = \frac{2}{X}\frac{Y'}{Y} \tag{126}$$

$$(\Theta/c) = F\left(\frac{\dot{X}/c}{X} + 2\,\frac{\dot{Y}/c}{Y}\right) \tag{127}$$

$$(\sigma/c) = \frac{F}{\sqrt{3}}\left(\frac{\dot{X}/c}{X} - \frac{\dot{Y}/c}{Y}\right) \tag{128}$$

$$(\dot{u}/c^2) = -\frac{1}{X}\frac{F'}{F}\,. \tag{129}$$

Eqs. (108) and (112) can then be integrated to give

$$K = \frac{C_1}{Y^2}\,, \tag{130}$$

with $C_1$ an integration constant. Next, from Eq. (101), we obtain the differential expression

$${}^3\!R = -\frac{2}{X^2}\left[\,2\,\frac{Y''}{Y} - 2\,\frac{X'}{X}\frac{Y'}{Y} + \left(\frac{Y'}{Y}\right)^2\,\right] + 2\,\frac{C_1}{Y^2}\,. \tag{131}$$

In integrating the field equations Eq. (101) is commonly used in place of Eq. (104).

Exact solutions to the dynamical equations within this LRS class have been discussed by Kramer et al, chapter 14.2 [1], for the spherically symmetric case ($K > 0$).

By employing the identity (47), from the set of equations (116) - (124) we can derive the following useful relations, which show the non-linear growth of spatial inhomogeneities within this LRS class (cf. Refs. [22, 23]):

$$[\,\mu'\,]^{\cdot}/c = -\tfrac{2}{3}\,[\,2\,(\Theta/c) + \sqrt{3}\,(\sigma/c)\,]\,\mu' - (\mu+p)\,(\Theta/c)' \tag{132}$$

$$\begin{aligned}[\,(\Theta/c)'\,]^{\cdot}/c = {} & -[\,(\Theta/c) + 2\sqrt{3}\,(\sigma/c)\,]\,(\Theta/c)' + (\dot{u}/c^2)'' + a\,(\dot{u}/c^2)' + 3\,(\dot{u}/c^2)\,(\dot{u}/c^2)' \\ & + 6\,a\,(\sigma/c)^2 - (\dot{u}/c^2)\,\Big[\,\tfrac{1}{9}\,(\Theta/c)^2 - \tfrac{2}{3\sqrt{3}}\,(\Theta/c)\,(\sigma/c) + \tfrac{10}{3}\,(\sigma/c)^2 + \tfrac{2}{\sqrt{3}}\,E \\ & + \tfrac{a^2}{2} - \tfrac{2}{3}\,\tfrac{4\pi G}{c^4}\,\mu - \tfrac{4\pi G}{c^4}\,(\mu+p)\left(\tfrac{\partial p}{\partial\mu}\right)^{-1} - a\,(\dot{u}/c^2) - (\dot{u}/c^2)^2\,\Big]\end{aligned} \tag{133}$$

$$\begin{aligned}[\,(\sigma/c)'\,]^{\cdot}/c = {} & -\tfrac{1}{\sqrt{3}}\,[\,(\Theta/c) + 2\sqrt{3}\,(\sigma/c)\,]\,(\Theta/c)' + \tfrac{1}{\sqrt{3}}\,(\dot{u}/c^2)'' - \tfrac{a}{2\sqrt{3}}\,(\dot{u}/c^2)' \\ & + \sqrt{3}\,(\dot{u}/c^2)\,(\dot{u}/c^2)' + 2\sqrt{3}\,a\,(\sigma/c)^2 + \tfrac{3}{2}\,a\,(\Theta/c)\,(\sigma/c) + \tfrac{3}{2}\,a\,E \\ & - \tfrac{1}{\sqrt{3}}\,(\dot{u}/c^2)\,\Big[\,\tfrac{1}{9}\,(\Theta/c)^2 + \tfrac{7}{3\sqrt{3}}\,(\Theta/c)\,(\sigma/c) + \tfrac{1}{3}\,(\sigma/c)^2 + \tfrac{2}{\sqrt{3}}\,E \\ & - \tfrac{a^2}{4} - \tfrac{2}{3}\,\tfrac{4\pi G}{c^4}\,\mu - \tfrac{4\pi G}{c^4}\,(\mu+p)\left(\tfrac{\partial p}{\partial\mu}\right)^{-1} + \tfrac{a}{2}\,(\dot{u}/c^2) - (\dot{u}/c^2)^2\,\Big]\end{aligned} \tag{134}$$

$$\begin{aligned}[\,E'\,]^{\cdot}/c = {} & -\tfrac{2}{3}\,[\,2\,(\Theta/c) + \sqrt{3}\,(\sigma/c)\,]\,\tfrac{4\pi G}{\sqrt{3}c^4}\,\mu' - \tfrac{4\pi G}{\sqrt{3}c^4}\,(\mu+p)\,(\Theta/c)' \\ & + \tfrac{3}{2}\,\tfrac{4\pi G}{c^4}\,(\mu+p)\,a\,(\sigma/c) + [\,2\,a - (\dot{u}/c^2)\,]\,[\,(\Theta/c) - \sqrt{3}\,(\sigma/c)\,]\,E\end{aligned} \tag{135}$$

$$\begin{aligned}[\,a'\,]^{\cdot}/c = {} & \tfrac{2}{3}\,[\,(\Theta/c) - \sqrt{3}\,(\sigma/c)\,]\,(\dot{u}/c^2)' - \tfrac{4}{27}\,(\Theta/c)^3 - \tfrac{2}{3\sqrt{3}}\,(\Theta/c)^2\,(\sigma/c) \\ & + \tfrac{2}{3}\,(\Theta/c)\,(\sigma/c)^2 + \tfrac{4}{3\sqrt{3}}\,(\sigma/c)^3 + \tfrac{a^2}{3}\,[\,(\Theta/c) - \sqrt{3}\,(\sigma/c)\,]\end{aligned}$$

$$
\begin{aligned}
& + \tfrac{2}{3\sqrt{3}} \, [\, 2\,(\Theta/c) + \sqrt{3}\,(\sigma/c) \,] \left[\, E + 2\,\tfrac{4\pi G}{\sqrt{3}c^4}\,\mu \,\right] \\
& - (\dot{u}/c^2) \left[\, \tfrac{a}{3}\,(\Theta/c) - \tfrac{4}{\sqrt{3}}\,a\,(\sigma/c) - \tfrac{2}{3}\,(\Theta/c)\,(\dot{u}/c^2) + \tfrac{2}{\sqrt{3}}\,(\sigma/c)\,(\dot{u}/c^2) \,\right] \qquad (136)
\end{aligned}
$$

$$
\begin{aligned}
[\,{}^3R'\,]^{\cdot}/c \;=\; & -4\,[\,(\Theta/c) + \sqrt{3}\left(1 + \tfrac{\partial p}{\partial \mu}\right)(\sigma/c)\,]\,\tfrac{4\pi G}{c^4}\,\mu' \\
& + \left[\, \tfrac{16}{9}\,(\Theta/c)^2 + \tfrac{4}{3\sqrt{3}}\,(\Theta/c)\,(\sigma/c) - \tfrac{20}{3}\,(\sigma/c)^2 - 2\,a\,(\dot{u}/c^2) - \tfrac{4}{\sqrt{3}}\,E - \tfrac{8}{3}\,\tfrac{4\pi G}{c^4}\,\mu \,\right](\Theta/c)' \\
& + 2\,a\,(\sigma/c)\left[\,(\Theta/c)\,(\sigma/c) + 5\sqrt{3}\,(\sigma/c)^2 + 6\,E\,\right] \\
& - \tfrac{4}{3}\,[\,(\Theta/c) - \sqrt{3}\,(\sigma/c)\,]\,(\dot{u}/c^2)'' - \tfrac{4}{3}\,a\,[\,(\Theta/c) + 2\sqrt{3}\,(\sigma/c)\,]\,(\dot{u}/c^2)' \\
& - 4\,[\,(\Theta/c) - \sqrt{3}\,(\sigma/c)\,]\,(\dot{u}/c^2)\,(\dot{u}/c^2)' - (\dot{u}/c^2)\Big[ -\tfrac{4}{27}\,(\Theta/c)^3 + \tfrac{4}{3\sqrt{3}}\,(\Theta/c)^2\,(\sigma/c) \\
& - \tfrac{4}{3}\,(\Theta/c)\,(\sigma/c)^2 + \tfrac{4}{3\sqrt{3}}\,(\sigma/c)^3 - \tfrac{8}{3\sqrt{3}}\,[\,(\Theta/c) - \sqrt{3}\,(\sigma/c)\,]\,E \\
& + \tfrac{4}{3}\,[\,(\Theta/c) + 2\sqrt{3}\,(\sigma/c)\,]\,a\,(\dot{u}/c^2) + \tfrac{4}{3}\,\tfrac{4\pi G}{\sqrt{3}c^4}\,(7\mu + 9p)\,(\sigma/c) + \tfrac{8}{9}\,\tfrac{4\pi G}{c^4}\,\mu\,(\Theta/c) \\
& - \tfrac{2}{3}\,a^2\,[\,(\Theta/c) + 2\sqrt{3}\,(\sigma/c)\,] + \tfrac{4}{3}\,[\,(\Theta/c) - \sqrt{3}\,(\sigma/c)\,]\,(\dot{u}/c^2)^2 \Big] \;. \qquad (137)
\end{aligned}
$$

In scenarios of the formation of structure on cosmological distance scales it is well-established practice to study the evolution of *linearised* perturbations around an assumed FLRW background spacetime geometry. In their FLRW-linearised form the set of equations (132) - (137) simplifies significantly and could be used for example to investigate covariant and gauge-invariant spherically symmetric perturbations of an FLRW spacetime [22, 23]. The description of scalar (total energy density) perturbations, which are the only ones contributing at linear order, would be entirely covered by only two equations from this set, for example Eqs. (132) and (133), and the equations underlying the evolution of the FLRW background spacetime geometry (see subsection 7.2.1 below).

## 6.1 Spatially inhomogeneous LRS dust models

With the condition $p = 0 \Leftrightarrow (\dot{u}/c^2) = 0$ imposed, the cosmological models within this LRS subclass are the Lemaître–Tolman–Bondi spherically symmetric solutions [24, 25, 26] and their generalisations to spacelike 2-surfaces with vanishing or negative Gaußian curvature scalar $K$[5]. These spatially inhomogeneous dust models belong to the so-called "silent" class recently discussed by Bruni et al [27]. The relevant equations, which we obtain from specialisation of the set (116) - (124), are

$$
\begin{aligned}
(\Theta/c)^{\cdot}/c &= -\tfrac{1}{3}\,(\Theta/c)^2 - 2\,(\sigma/c)^2 - \tfrac{4\pi G}{c^4}\,\mu && (138) \\
(\sigma/c)^{\cdot}/c &= -\tfrac{1}{\sqrt{3}}\,(\sigma/c)^2 - \tfrac{2}{3}\,(\Theta/c)\,(\sigma/c) - E && (139) \\
\dot{E}/c &= -\tfrac{4\pi G}{c^4}\,\mu\,(\sigma/c) + \sqrt{3}\,E\,(\sigma/c) - (\Theta/c)\,E && (140) \\
\dot{\mu}/c &= -\mu\,(\Theta/c) && (141) \\
\dot{a}/c &= -\tfrac{a}{3}\,(\Theta/c) + \tfrac{a}{\sqrt{3}}\,(\sigma/c) && (142) \\
(\sigma/c)' &= \tfrac{1}{\sqrt{3}}\,(\Theta/c)' - \tfrac{3}{2}\,a\,(\sigma/c) && (143) \\
E' &= -\tfrac{3}{2}\,a\,E + \tfrac{4\pi G}{\sqrt{3}c^4}\,\mu' && (144) \\
a' &= \tfrac{2}{9}\,(\Theta/c)^2 + \tfrac{2}{3\sqrt{3}}\,(\Theta/c)\,(\sigma/c) - \tfrac{4}{3}\,(\sigma/c)^2 - \tfrac{2}{\sqrt{3}}\,E - \tfrac{a^2}{2} - \tfrac{4}{3}\,\tfrac{4\pi G}{c^4}\,\mu \;. && (145)
\end{aligned}
$$

**Algorithm:** The *initial data*, which can be specified *freely* on a spacelike 3-surface orthogonal to the matter flow, are the values of $\mu$ and $(\Theta/c)$, while $(\sigma/c)$, $E$ and $a$ can be given at a point and their spatial distribution is determined by Eqs. (143), (144) and (145) respectively. Then all time derivatives are known and $K$ is given by Eq. (115).

Exact solutions to the dynamical equations within this LRS subclass have been discussed by Lemaître [24], Tolman [25], Bondi [26], and Kramer et al, chapter 13.5 [1], for the spherically symmetric case ($K > 0$). Ellis [5] discussed the more general cases. For dust, in the local comoving coordinates of Eq. (125), we have from Eq. (129) that $F = 1$. The set of equations (132) - (137) describing the non-linear growth of spatial inhomogeneities simplifies considerably and reads:

$$\begin{aligned}
[\,\mu'\,]^{\cdot}/c &= -\tfrac{2}{3}\,[\,2\,(\Theta/c)+\sqrt{3}\,(\sigma/c)\,]\,\mu' - \mu\,(\Theta/c)' && (146)\\
\left[\,(\Theta/c)'\,\right]^{\cdot}/c &= -\tfrac{4\pi G}{c^4}\,\mu' - [\,(\Theta/c)+2\sqrt{3}\,(\sigma/c)\,]\,(\Theta/c)' + 6\,a\,(\sigma/c)^2 && (147)\\
\left[\,(\sigma/c)'\,\right]^{\cdot}/c &= -\tfrac{4\pi G}{\sqrt{3}c^4}\,\mu' - \tfrac{1}{\sqrt{3}}\,[\,(\Theta/c)+2\sqrt{3}\,(\sigma/c)\,]\,(\Theta/c)' + 2\sqrt{3}\,a\,(\sigma/c)^2 \\
&\quad + \tfrac{3}{2}\,a\,(\Theta/c)\,(\sigma/c) + \tfrac{3}{2}\,a\,E && (148)\\
[\,E'\,]^{\cdot}/c &= -\tfrac{2}{3}\,[\,2\,(\Theta/c)+\sqrt{3}\,(\sigma/c)\,]\,\tfrac{4\pi G}{\sqrt{3}c^4}\,\mu' - \tfrac{4\pi G}{\sqrt{3}c^4}\,\mu\,(\Theta/c)' + \tfrac{3}{2}\,\tfrac{4\pi G}{c^4}\,\mu\,a\,(\sigma/c) \\
&\quad + 2\,a\,E\,[\,(\Theta/c)-\sqrt{3}\,(\sigma/c)\,] && (149)\\
[\,a'\,]^{\cdot}/c &= -\tfrac{4}{27}\,(\Theta/c)^3 - \tfrac{2}{3\sqrt{3}}\,(\Theta/c)^2\,(\sigma/c) + \tfrac{2}{3}\,(\Theta/c)\,(\sigma/c)^2 + \tfrac{4}{3\sqrt{3}}\,(\sigma/c)^3 \\
&\quad + \tfrac{a^2}{3}\,[\,(\Theta/c)-\sqrt{3}\,(\sigma/c)\,] + \tfrac{2}{3\sqrt{3}}\,[\,2\,(\Theta/c)+\sqrt{3}\,(\sigma/c)\,]\,\left[\,E + 2\,\tfrac{4\pi G}{\sqrt{3}c^4}\,\mu\,\right] && (150)\\
[\,{}^3\!R'\,]^{\cdot}/c &= -4\,[\,(\Theta/c)+\sqrt{3}\,(\sigma/c)\,]\,\tfrac{4\pi G}{c^4}\,\mu' \\
&\quad + \left[\,\tfrac{16}{9}\,(\Theta/c)^2 + \tfrac{4}{3\sqrt{3}}\,(\Theta/c)\,(\sigma/c) - \tfrac{20}{3}\,(\sigma/c)^2 - \tfrac{4}{\sqrt{3}}\,E - \tfrac{8}{3}\,\tfrac{4\pi G}{c^4}\,\mu\,\right](\Theta/c)' \\
&\quad + 2\,a\,(\sigma/c)\left[\,(\Theta/c)\,(\sigma/c) + 5\sqrt{3}\,(\sigma/c)^2 + 6\,E\,\right] . && (151)
\end{aligned}$$

The linearised set of equations corresponding to [22] easily follows.

## 6.2 $(\sigma/c) = 0$: The shearfree subcase

An interesting subcase within LRS class II arises, if we demand that $(\sigma/c) = 0$. Then we immediately obtain from Eq. (117) that the magnitude of the "electric part" of the conformal Weyl curvature tensor is given by

$$E = \tfrac{1}{\sqrt{3}}\,(\dot{u}/c^2)' - \tfrac{a}{2\sqrt{3}}\,(\dot{u}/c^2) + \tfrac{1}{\sqrt{3}}\,(\dot{u}/c^2)^2 \ . \tag{152}$$

Hence, in order to obtain a solution with $E \neq 0$, the fluid acceleration has to be non-zero, and we also assume a non-zero rate of expansion of the matter fluid. (The case with $(\Theta/c) = 0$ will be treated in subsection 7.1 below.) Next, from the constraint (121) it follows that

$$(\Theta/c)' = 0 \ , \tag{153}$$

that is, the spatial distribution of the rate of expansion is homogeneous and thus constant on the spacelike 3-surfaces orthogonal to $u^i/c$. With Eqs. (152) and (110) the conditions deriving from Eqs. (133) and (134) for $(\sigma/c) = 0$ are equivalent: they provide the expression

$$\begin{aligned}
(\dot{u}/c^2)'' &= \tfrac{4\pi G}{c^4}\,\mu' - a\,(\dot{u}/c^2)' - \tfrac{7}{3}\,(\dot{u}/c^2)\,(\dot{u}/c^2)' \\
&\quad + (\dot{u}/c^2)\left[\,\tfrac{1}{9}\,(\Theta/c)^2 + \tfrac{a^2}{2} - \tfrac{2}{3}\,\tfrac{4\pi G}{c^4}\,\mu - \tfrac{4}{3}\,a\,(\dot{u}/c^2) - \tfrac{1}{3}\,(\dot{u}/c^2)^2\,\right] , && (154)
\end{aligned}$$

which then ensures that the constraint (122) is solved identically. Using relation (110), Eq. (154) is a constraint equation for the spatial distribution of the total energy density $\mu$ of third order, which is coupled to the constraint equation (124), giving the spatial distribution of the spatial divergence $a$.

In the local comoving coordinates of Eq. (125) we have that

$$(\sigma/c) = 0 \Rightarrow F\,\frac{\dot{X}/c}{X} = F\,\frac{\dot{Y}/c}{Y} \Rightarrow (\Theta/c) = 3\,F\,\frac{\dot{X}/c}{X} \ . \tag{155}$$

Then we can immediately integrate Eq. (132) to find

$$\mu' = \frac{C_2}{X^4} \ , \tag{156}$$

where $C_2$ denotes an integration constant. Using this result and Eqs. (110) and (129), an expression for $\frac{\partial p}{\partial \mu}$ is established.

Exact solutions to the dynamical equations within this LRS subclass have been discussed in Kramer et al, chapter 14.2.3 [1], for the spherically symmetric case $(K > 0)$ and $p = p(\mu)$.

It would be interesting to investigate, whether a solution with $(\Theta/c)^{\cdot}/c = 0 \Rightarrow (\Theta/c) = \text{const}$, $(\mu + 3p) > 0$ and *negative* deceleration parameter $q$ exists within this subclass of LRS class II, which is consistent. A solution of this kind would provide a very special spatially inhomogeneous inflationary cosmological model.

### 6.3 $a = 0$: The non-diverging subcase

If $a = 0$, the normals to the spacelike 2-surfaces spanned by the isometry group are non-diverging. The conditions for this dynamical restriction to be consistent are the following: for $a = 0$ Eq. (120) demands that

$$0 = [\,(\Theta/c) - \sqrt{3}\,(\sigma/c)\,]\,\left(\dot{u}/c^2\right) \ , \tag{157}$$

while Eq. (124) gives $E$ algebraically. If (i) $(\Theta/c) = \sqrt{3}\,(\sigma/c)$, inserting $E$ into the constraint (122) requires that $\mu' = 0$, while Eqs. (117) and (118) give $(\mu + p) = 0$ and $(\mu + p)\,(\Theta/c) = 0$ respectively. Consequently, this case can be discarded. If on the other hand (ii) $\left(\dot{u}/c^2\right) = 0$, for $f' \neq 0$ we need to have $p = 0$. Then from Eq. (144) we get the condition

$$(\Theta/c)' = 3\,[\,(\Theta/c) - \sqrt{3}\,(\sigma/c)\,]^{-1}\,\tfrac{4\pi G}{c^4}\,\mu' \ , \tag{158}$$

and the covariant time derivatives along the matter fluid flow lines of the constraints (143), (144) and (145) vanish identically. That is, only dust provides an appropriate matter source for spatially inhomogeneous models of LRS class II with non-diverging isotropy generator. These models are thus a further specialisation of the "silent" class [27] discussed in subsection 6.1.

In conclusion, the generically spatially inhomogeneous LRS class II constitutes the largest class of solutions of perfect fluid spacetime geometries with equation of state $p = p(\mu)$. The models are in general time-dependent, and in the past they have been prime candidates for the theoretical and numerical description of star and galaxy formation processes as well as supernova explosions (see Refs. [9], [10], [14] and [15]). Due to their highly idealised spacetime symmetry properties the relevant dynamical equations become relatively simple and easily tractable. However, as outlined in the introduction, an isentropic matter fluid flow, resulting from a barotropic equation of state, will in general be too restrictive to realistically model for example explosions of stars during the late stages of their evolution.

In the following section we discuss those subcases of cosmological models within LRS class II, which contain an additional (timelike or spacelike) Killing symmetry, apart from the multiply-transitive $G_3$ isometry group.

## 7 Higher Symmetry Subcases: Hypersurface Homogeneous Models in LRS class II

Given generic initial data in the spherically symmetric family, we obtain general spherically symmetric inhomogeneous solutions, and their plane symmetric and hyperbolic generalisations, as discussed in the last section. Now we consider those cases where the initial conditions chosen result in higher symmetry solutions. By this we mean in particular that through such a special choice a further Killing spacetime symmetry shall arise. As a preliminary we note that if we have for example spatial homogeneity, then necessarily any covariantly defined scalar $f$ is functionally dependent on say the total energy density $\mu$ [28, 29]. That is

$$f = f(\mu) \tag{159}$$

and so

$$\nabla_i f = \frac{df(\mu)}{d\mu}\,\nabla_i \mu \ , \tag{160}$$

which implies the two relations

$$F_i := h^j{}_i\,\nabla_j f = \frac{df(\mu)}{d\mu}\,h^j{}_i\,\nabla_j \mu := \frac{df(\mu)}{d\mu}\,X_i \ , \tag{161}$$

and

$$\dot{f}/c = \frac{df(\mu)}{d\mu}\,\dot{\mu}/c\;. \tag{162}$$

Similar relations will hold for each pair of scalars, like for example $\{(\Theta/c), f\}$. Before discussing the spatially homogeneous subcases within LRS class II in detail, we first want to turn our attention to the situation where the special choice of initial data gives rise to timelike 3-surfaces of homogeneity.

## 7.1 $\dot{f}/c = 0$: The static subcase

Imposing the geometrical condition $\dot{f}/c = 0$ on the set of equations (116) - (124), that is, allowing for the additional timelike Killing vector field $\partial_{ct}$ such that the group of isometries is a $G_4$ multiply-transitive on timelike 3-surfaces, for $(\mu + p) > 0$, $E \neq 0$, we immediately get from Eqs. (119) and (118) that $(\Theta/c) = 0 = (\sigma/c)$. Then by combining Eqs. (116) and (117) we obtain for the magnitude $E$

$$E = \tfrac{4\pi G}{\sqrt{3}c^4}\,(\mu + 3p) - \tfrac{\sqrt{3}}{2}\,a\left(\dot{u}/c^2\right)\;, \tag{163}$$

which, when substituted into Eq. (115), yields for the (constant) Gaußian curvature of the spacelike 2-D symmetry orbits of the $G_3$ subgroup

$$K = \tfrac{a^2}{4} + a\left(\dot{u}/c^2\right) - 2\,\tfrac{4\pi G}{c^4}\,p\;. \tag{164}$$

In principle $K$ can be positive, negative, or zero. Finally, the remaining equations of the set (116) - (124) lead to the set

$$\left(\dot{u}/c^2\right)' = -a\left(\dot{u}/c^2\right) - \left(\dot{u}/c^2\right)^2 + \tfrac{4\pi G}{c^4}\,(\mu + 3p) \tag{165}$$

$$\left(\dot{u}/c^2\right) = -\tfrac{\partial p}{\partial \mu}\,\mu'\,/\,(\mu + p) \;=\; -p'\,/\,(\mu + p) \tag{166}$$

$$a' = -\tfrac{a^2}{2} + a\left(\dot{u}/c^2\right) - 2\,\tfrac{4\pi G}{c^4}\,(\mu + p)\;. \tag{167}$$

With Eq. (163) the constraint equation (122) is identically satisfied.

In the local comoving coordinates of Eq. (125) we have that $X = X(x)$, $Y = Y(x)$, $F = F(x)$. In the case of modeling static, spherically symmetric matter configurations (where $K > 0$), however, it is more convenient to use Schwarzschild coordinates such that the line element assumes the form [15]

$$ds^2 = e^{\lambda(r)}\,dr^2 + r^2\left(d\vartheta^2 + \sin^2\vartheta\,d\varphi^2\right) - e^{\nu(r)}\,d(ct)^2\;, \tag{168}$$

and the fluid 4-velocity is given by $u^i/c = e^{-\nu/2}\,\delta^i{}_4$. Upon combination with the Einstein field equations this leads to [15]

$$\left(\dot{u}/c^2\right) = \left[\,1 - \tfrac{2\,m(r)}{r}\,\right]^{-1}\,\tfrac{1}{r^2}\,\left[\,m(r) + \tfrac{4\pi G}{c^4}\,r^3\,p\,\right]\;, \tag{169}$$

where $m(r) := \frac{4\pi G}{c^4}\int_0^r \mu(x)\,x^2\,dx$. Inserting (169) into Eq. (166) then yields the Tolman–Oppenheimer–Volkoff structure equation, which describes the interior spacetime geometry of static, spherically symmetric relativistic stars [9, 10]. If we use expression (169) in Eqs. (165) and (167), we have to solve a coupled set of two ordinary differential equations for the two functions $\mu(r)$ and $a(r)$, provided an equation of state $p = p(\mu)$ has been specified.

In the remaining two subsections we want to address the spatially homogeneous subcases within LRS class II, which generally occur in two different ways.

## 7.2 Orthogonal spatially homogeneous models (OSH)

This is the higher symmetry family in which there is an extra symmetry acting on spacelike 3-surfaces orthogonal to $u^i/c$, and with shear. Consequently, additional to $(\omega/c) = 0$, we have $f' = 0$ for all scalars, which necessarily implies from Eq. (123) that $\left(\dot{u}/c^2\right) = 0$, and also from Eqs. (121) and (122) that $a = 0$. Thus with $f' = 0$, Eq. (161) is identically satisfied because we have that

$$X_i = 0\;, \qquad F_i = 0\;. \tag{170}$$

Equation (162) will follow from the dynamics of the system. It does not tell us anything about spatial homogeneity.

The 3-Ricci tensor for OSH models of this subclass is given from Eq. (100) by

$$ {}^3R_{ij} = \tfrac{2}{3}\,K\,(h_{ij} - e_{ij}) = K\,p_{ij} \ , \tag{171} $$

while the generalised Friedmann equation is

$$ {}^3R = 2\,K = 4\,\tfrac{4\pi G}{c^4}\,\mu - \tfrac{2}{3}\,(\Theta/c)^2 + 2\,(\sigma/c)^2 \ . \tag{172} $$

The evolution of ${}^3R$ is governed by the equation

$$ [\ {}^3R\ ]\,\dot{}\,/c = -\tfrac{1}{3}\ [\ 2\,(\Theta/c) + \sqrt{3}\,(\sigma/c)\ ]\ {}^3R + 2\sqrt{3}\,K\,(\sigma/c) \ , \tag{173} $$

and from Eq. (124) it follows that

$$ E = \tfrac{1}{3\sqrt{3}}\,(\Theta/c)^2 + \tfrac{1}{3}\,(\Theta/c)\,(\sigma/c) - \tfrac{2}{\sqrt{3}}\,(\sigma/c)^2 - 2\,\tfrac{4\pi G}{\sqrt{3}c^4}\,\mu \ . \tag{174} $$

The dynamical equations (116) - (124) reduce to the set

$$ \begin{aligned} (\Theta/c)\dot{}/c &= -\tfrac{1}{3}\,(\Theta/c)^2 - 2\,(\sigma/c)^2 - \tfrac{4\pi G}{c^4}\,(\mu + 3p) & (175)\\ (\sigma/c)\dot{}/c &= -\tfrac{1}{3\sqrt{3}}\,(\Theta/c)^2 - (\Theta/c)\,(\sigma/c) + \tfrac{1}{\sqrt{3}}\,(\sigma/c)^2 + 2\,\tfrac{4\pi G}{\sqrt{3}c^4}\,\mu & (176)\\ \dot{\mu}/c &= -\,(\mu + p)\,(\Theta/c) \ . & (177) \end{aligned} $$

The isometry group underlying OSH models within this subclass can be of Bianchi Type–I/VII$_0$ (where the group orbits have isotropic 3-Ricci curvature with ${}^3R = 0 = 2\,K$), or of a special case of Bianchi Type-III (with $K < 0$) [7], or instead it can be the Kantowski–Sachs (KS) case [30] (with $K > 0$), which is the *unique* case where the group of isometries $G_4$ multiply-transitive on spacelike 3-surfaces orthogonal to $u^i/c$ does *not* contain a simply-transitive $G_3$ subgroup [4, 31].

In the local comoving coordinates of Eq. (125) we have that $X = X(ct)$, $Y = Y(ct)$, $F = 1$, and the solution to Eq. (173) is given by

$$ {}^3R = 2\,\frac{C_1}{Y^2} + \frac{C_2}{X\,Y} \ , \tag{178} $$

where $C_1$ and $C_2$ are integration constants. Thus Eq. (172) constitutes a first integral and is generally used in place of Eq. (175).

### 7.2.1 FLRW

Here we have that $(\dot{u}/c^2) = 0 \Leftrightarrow (\sigma/c) = 0 \Leftrightarrow E = 0$. This is a subcase of the previous one, but it *no longer* follows that necessarily $a = 0$, $a' = 0$, as this is the exceptional case where the spacelike unit vector field $e^i$ is *not* uniquely defined. As mentioned in subsection 5.3 there is *no* covariant feature that picks out a preferred spatial direction.

With the condition $a' = -\,2\,K$ required to obtain spacelike 3-surfaces orthogonal to $u^i/c$, which have isotropic curvature (cf. Eq. (100)), the Friedmann equation is given by

$$ {}^3R = 6\,\left[\ K - \tfrac{a^2}{4}\ \right] = 4\,\tfrac{4\pi G}{c^4}\,\mu - \tfrac{2}{3}\,(\Theta/c)^2 \ . \tag{179} $$

The evolution and constraint equations for the non-zero variables are

$$ \begin{aligned} (\Theta/c)\dot{}/c &= -\tfrac{1}{3}\,(\Theta/c)^2 - \tfrac{4\pi G}{c^4}\,(\mu + 3p) & (180)\\ \dot{\mu}/c &= -\,(\mu + p)\,(\Theta/c) & (181)\\ \dot{K}/c &= -\tfrac{2}{3}\,K\,(\Theta/c) & (182)\\ \dot{a}/c &= -\tfrac{a}{3}\,(\Theta/c) & (183)\\ [\ {}^3R\ ]\,\dot{}\,/c &= -\tfrac{2}{3}\,(\Theta/c)\ {}^3R & (184)\\ K' &= -\,a\,K \ . & (185) \end{aligned} $$

However, once the sign of ${}^3R$ was fixed, where either ${}^3R > 0$, ${}^3R = 0$ or ${}^3R < 0$, only two of these equations are essential, which are commonly chosen to be Eq. (179) and Eq. (181). In the local comoving coordinates of Eq. (125) we have that $X = X(ct)$, $Y = X(ct)\,Z(x)$, $F = 1$, and Eq. (184) can be integrated to give

$$ {}^3R = \frac{C}{X^2} \ , \tag{186} $$

where $C$ is an integration constant.

The Einstein static model is the special FLRW case when $(\Theta/c) = 0$ and thus $1/6\,{}^3R = K - a^2/4 = 2/3\,\frac{4\pi G}{c^4}\,\mu > 0$. From Eq. (180) we then find that for consistency an equation of state of the form $p(\mu) = -1/3\,\mu =$ const is required, which can be interpreted as a dust model with a *positive* cosmological constant, $\Lambda > 0$, as was already discussed at the end of subsection 5.3.

#### 7.2.2 The dust subcases

Because the fluid flow is geodesic in any case, the dust subcases are simply distinguished by marginally simpler evolution equations: $p = 0$ in Eqs. (175) and (177).

### 7.3 Tilted spatially homogeneous LRS dust models

Here there exists an extra symmetry, but not in the 3-surfaces orthogonal to $u^i/c$. Instead, the 3-surfaces of homogeneity are *tilted* with respect to $u^i/c$, that is $u^i/c$ does *not* provide the unit normal to these 3-surfaces [8]. In the tilted case, with the cosmological conditions $(\Theta/c) \neq 0$, $(\mu + p) > 0$ imposed (which we will assume), we will have $X_i \neq 0$, $F_i \neq 0$. Then equation (161) implies that their spatial directions are parallel:

$$ \eta^{ijkl}\, X_j\, F_k\, u_l/c = 0 \ . \tag{187} $$

However from the LRS condition we have $X_i = \mu'\,e_i$ and $F_i = f'\,e_i$, so (187) will be trivially satisfied. Thus we also need the magnitude information. We don't know $f(\mu)$ (see (159), but we do know it must be the same in both equations, so we get a non-trivial magnitude relation by eliminating $df(\mu)/d\mu$ between the two equations (161) and (162), leading to

$$ \dot{\mu}/c\ F_i = \dot{f}/c\ X_i \qquad \Leftrightarrow \qquad \dot{\mu}/c\ f' = \dot{f}/c\ \mu' \ . \tag{188} $$

which is another way of saying that the surfaces of constant $\mu$ and $f$ must be the same. The corresponding set of equations to Eqs. (188) will hold for *every* pair of scalar invariants $f$, defined from the kinematic quantities, matter variables, Weyl curvature tensor, and *all* their covariant derivatives.

The key question in each case, related both to the Postlate of Uniformal Thermal Histories (PUTH) [32], and to the Equivalence Problem series of papers [12, 13], is *how few* such relations we have to impose before consistency conditions imply all the rest must be satisfied; and which variables should we choose for these relations. The weak conjecture is that $S_4 = \{\,\mu, (\Theta/c)\,, (\sigma/c)\,, E\,\}$ will suffice. The strong conjecture is that $S_2 = \{\,\mu, (\Theta/c)\,\}$ will suffice. However the latter may be disproved by the PUTH counter-examples.

It is known that the perfect fluid cases of this class are of Bianchi Type–V (see King and Ellis [8]), studied in some detail in [33] and [34]. However for reasons of simplicity, we here focus on the situation where the equation of state of the matter is that of dust, $p = 0 \Leftrightarrow (\dot{u}/c^2) = 0$. The exact solution to the field equations for this non-rotating tilted LRS SH dust model was given by Farnsworth [35]. It belongs to the "silent" class [27] of subsection 6.1. We assume the functional dependency relations $(\Theta/c) = (\Theta/c)[\mu]$, $(\sigma/c) = (\sigma/c)[\mu]$, $E = E[\mu]$, $a = a[\mu]$ and $K = K[\mu]$, and that all of these dynamical variables are *non-zero*. Then, using the set of dynamical equations (138) - (145), from Eq. (188) the following five consistency equations need to be satisfied:
First,

$$ \dot{\mu}/c\ \ (\Theta/c)' - (\Theta/c)\dot{}/c\ \mu' = 0 \ , \tag{189} $$

which can be solved to give

$$ (\Theta/c)' = \frac{\frac{4\pi G}{c^4}\,\mu + \frac{1}{3}\,(\Theta/c)^2 + 2\,(\sigma/c)^2}{\mu\ (\Theta/c)}\ \mu' \ . \tag{190} $$

Second,

$$\dot{\mu}/c\ \ (\sigma/c)' - (\sigma/c)\dot{}/c\ \mu' = 0\ , \tag{191}$$

which, on using Eq. (190), gives

$$\mu' = -\tfrac{3}{2}\ \frac{a\,\mu\,(\Theta/c)\,(\sigma/c)}{E - \frac{4\pi G}{\sqrt{3}c^4}\,\mu - \frac{1}{3\sqrt{3}}\,[\ (\Theta/c) - \sqrt{3}\,(\sigma/c)\ ]^2}\ . \tag{192}$$

Third,

$$\dot{\mu}/c\ E' - \dot{E}/c\ \mu' = 0\ , \tag{193}$$

which with Eq. (192) leads to the algebraic relation

$$\tfrac{4\pi G}{c^4}\,\mu = \sqrt{3}\ \frac{E - \frac{1}{3\sqrt{3}}\,[\ (\Theta/c) - \sqrt{3}\,(\sigma/c)\ ]\,[\ (\Theta/c) + 2\sqrt{3}\,(\sigma/c)\ ]}{E - [\ (\Theta/c) - \sqrt{3}\,(\sigma/c)\ ]\,(\sigma/c)}\ E\ . \tag{194}$$

Fourth, the condition

$$\dot{\mu}/c\ a' - \dot{a}/c\ \mu' = 0\ , \tag{195}$$

which, after inserting from Eqs. (190), (192) and (194), leads to a higher-order multivariate polynomial equation in the variables $(\Theta/c)$, $(\sigma/c)$, $E$ and $a$, which is quadratic in $a$, and finally the relation

$$\dot{\mu}/c\ K' - \dot{K}/c\ \mu' = 0\ , \tag{196}$$

which, after inserting from Eqs. (190), (192) and (194), yields a second higher-order multivariate polynomial equation in the variables $(\Theta/c)$, $(\sigma/c)$, $E$, that however does *not* contain the variable $a$.

The next step to take is to consider the covariant time derivatives along the matter fluid flow lines of the constraints (189), (191), (193), (195) and (196), and work out the consequences which establish that these be preserved in time. Those of the relations for the pairs $\{\,\mu, (\Theta/c)\,\}$, Eq. (189), $\{\,\mu, (\sigma/c)\,\}$, Eq. (191), and $\{\,\mu, E\,\}$, (193), vanish identically, when Eqs. (190), (192) and (194) are employed, while the relations for the pairs $\{\,\mu, a\,\}$, Eq. (195), and $\{\,\mu, K\,\}$, Eq. (196), give two further higher-order multivariate polynomial equations, the first again quadratic in $a$, while the second again does *not* contain $a$. Combining them with the two obtained from the functional dependency relations (195) and (196), we have to solve a set of four simultaneous non-linear algebraic equations in the variables $(\Theta/c)$, $(\sigma/c)$, $E$ and $a$. We tackle this problem by means of application of the `GRÖBNER` package of the algebraic computing system `REDUCE`. The solutions we find in this way are

$$(\Theta/c) = 0\ , \qquad (\sigma/c) = 0\ , \tag{197}$$

$$E = \tfrac{2}{3\sqrt{3}}\,(\Theta/c)^2\ , \qquad (\sigma/c) = \tfrac{1}{\sqrt{3}}\,(\Theta/c)\ , \tag{198}$$

$$E = -\tfrac{2}{3\sqrt{3}}\,(\Theta/c)^2\ , \qquad (\sigma/c) = -\tfrac{1}{\sqrt{3}}\,(\Theta/c)\ , \tag{199}$$

$$E = \tfrac{1}{3\sqrt{6}}\,(\Theta/c)^2\ , \qquad (\sigma/c) = \tfrac{1}{\sqrt{6}}\,(\Theta/c)\ , \tag{200}$$

$$E = -\tfrac{1}{3\sqrt{6}}\,(\Theta/c)^2\ , \qquad (\sigma/c) = -\tfrac{1}{\sqrt{6}}\,(\Theta/c)\ . \tag{201}$$

However, all of them have to be discarded, as they lead to consequences for $\mu$ and $\mu'$ which violate the initial assumptions. The failure of solution (197) is obvious, while (198) with (194) gives a zero denominator in Eq. (192), and solutions (199) - (201) lead to $\mu = 0$ from Eq. (194).

Thus, as one can show that the alternative assumptions $a = 0$, $K \neq 0$ and $a = 0$, $K = 0$ also lead to contradictions, we conclude that for consistency Eqs. (189) - (196) have to be solved for $K = 0$, $a \neq 0$ instead (see also [31]). That is, from Eq. (103) we have

$$E = \tfrac{4\pi G}{\sqrt{3}c^4}\,\mu + \tfrac{\sqrt{3}}{8}\,a^2 - \tfrac{1}{6\sqrt{3}}\ [\ (\Theta/c) - \sqrt{3}\,(\sigma/c)\ ]^2\ , \tag{202}$$

which simplifies our considerations significantly. With (202) the evolution and constraint equations (140) and (144) are identically satisfied. Now Eq. (190) is still valid, whereas from Eq. (191) we get

$$\mu' = \frac{3\sqrt{3}\,a\,\mu\ (\Theta/c)\,(\sigma/c)}{[\ (\Theta/c) - \sqrt{3}\,(\sigma/c)\ ]^2 - \frac{3}{4}\,a^2}\ , \tag{203}$$

which shows that for $a = 0$ we necessarily have $\mu' = 0$, and then successively $f' = 0$ for all scalar invariants $f$. With Eqs. (190) and (203) inserted, Eq. (195) yields the algebraic expression

$$\tfrac{4\pi G}{c^4}\,\mu = \tfrac{1}{6}\left[\,(\Theta/c)^2 - 3\,(\sigma/c)^2 - \tfrac{3}{4}\,a^2\,\right]\frac{\left[\,(\Theta/c) - \sqrt{3}\,(\sigma/c)\,\right]^2 - \tfrac{9}{4}\,a^2}{\left[\,(\Theta/c) - \sqrt{3}\,(\sigma/c)\,\right]^2 - \tfrac{3}{4}\,a^2}\,, \tag{204}$$

while Eq. (193) is identically solved. In principle, we can invert this (quartic) algebraic equation to determine the dynamical variable $a$ in terms of $\mu$, $(\Theta/c)$ and $(\sigma/c)$. Note that for $a = 0$ Eq. (204) reduces to Eq. (172) with $K = 0$. Note also that for $a^2 = 4/3\,[\,(\Theta/c) - \sqrt{3}\,(\sigma/c)\,]^2$, from Eq. (204) there occurs a singularity in the Ricci curvature tensor, and from Eq. (202) simultaneously in the Weyl curvature tensor. Equation (204) is compatible with Eq. (203), when Eqs. (143), (145), (190) and (202) are employed.

It remains to investigate the covariant time derivatives along the fluid flow lines of the constraints (189), (191), (193) and (195). We find that they vanish identically, after the expressions (190), (203) and (204) have been substituted, and thus consistency of our assumptions and results for the tilted SH dust subcase within the generically spatially inhomogeneous LRS class II has been established.

**Algorithm:** In the tilted SH dust subcase of LRS class II the *initial data*, which can be specified *freely* on a spacelike 3-surface orthogonal to the matter flow, are the values of $\mu$, $(\Theta/c)$ and $(\sigma/c)$ at a point. Then the values of $a$ and $E$ at that point follow from Eqs. (204) and (202) respectively. The spatial derivatives on the initial spacelike 3-surface orthogonal to $u^i/c$ of the non-zero dynamical variables are determined from Eqs. (203), (190), (143), (145) and (144), while their time derivatives come from Eqs. (138) - (142). $K = 0$, as demonstrated.

Applying the steps outlined in this subsection to the case with non-vanishing pressure and hence non-zero acceleration should lead to similar results. We leave this investigation open for future work.

# 8 Conclusion

All three distinct classes of spacetime geometries that arise for perfect fluids with equation of state $p = p(\mu)$ and LRS spacetime symmetries as well as their different subcases, defined through further either dynamical or geometrical restriction, have been discussed in detail in the relevant paragraphs of the last four sections. A short description of their various fields of application was given at the end of each section. Here we present a final compact survey of the different cases we have been dealing with in this paper and include references to the (sub)sections in which their discussion can be found.

This may be contrasted with the Table given at the end of [4], which characterises the subsets in a different way.

# Acknowledgements

HvE is supported by a grant from the Drapers' Society at QMW. GFRE acknowledges support from the FRD (South Africa). Throughout this work the computer algebra packages `REDUCE` and `CLASSI` have been valuable tools.

# References

[1] Kramer D, Stephani H, MacCallum M A H and Herlt E 1980 *Exact Solutions of Einstein's Field Equations* (Berlin: VEB Dt. Verlag d. Wissenschaften)

[2] Krasiński A 1993 *Physics in an Inhomogeneous Universe (A Review)* University of Cape Town Preprint 1993/10; To Appear, Cambridge University Press.

[3] Ehlers J 1961 "Beiträge zur relativistischen Mechanik kontinuierlicher Medien" *Akad. Wiss. Lit. Mainz, Abhandl. Math.-Nat. Kl.* **11**

Translation: Ehlers J 1993 "Contributions to the Relativistic Mechanics of Continuous Media" *GRG* **25** 1225

[4] Ellis G F R 1971 "Relativistic Cosmology" *General Relativity and Cosmology* Proceedings of the XLVII Enrico Fermi Summer School ed R K Sachs (New York: Academic Press)

[5] Ellis G F R 1967 "Dynamics of Pressure-Free Matter in General Relativity" *J. Math. Phys.* **8** 1171

[6] Stewart J M and Ellis G F R 1968 "Solutions of Einstein's Equations for a Fluid which Exhibits Local Rotational Symmetry" *J. Math. Phys.* **9** 1072

[7] Ellis G F R and MacCallum M A H 1969 "A Class of Homogeneous Cosmological Models" *Commun. Math. Phys.* **12** 108

[8] King A R and Ellis G F R 1973 "Tilted Homogeneous Cosmological Models" *Commun. Math. Phys.* **31** 209

[9] Misner C W, Thorne K S and Wheeler J A 1973 *Gravitation* (New York: Freeman & Co.)

[10] Shapiro S L and Teukolsky S A 1983 *Black Holes, White Dwarfs, and Neutron Stars: the Physics of Compact Objects* (New York: J Wiley & Sons)

[11] Wainwright J 1970 " A Class of Algebraically Special Perfect Fluid Space-Times" *Commun. Math. Phys.* **17** 42

[12] Bradley M and Karlhede A 1990 "On the Curvature Description of Gravitational Fields" *Class. Quantum Grav.* **7** 449

[13] MacCallum M A H and Skea J E F 1992 "`SHEEP`: A Computer Algebra System for General Relativity" *Proc. 1st Brazilian School on Computer Algebra* ed M J Rebouças (Oxford: Oxford University Press)

[14] Longair M S 1994 *High Energy Astrophysics (Vol 2): Stars, the Galaxy and the Interstellar Medium* (2nd Ed) (Cambridge: Cambridge University Press)

[15] Stephani H 1991 *Allgemeine Relativitätstheorie* (4. Auflage) (Berlin: Dt. Verlag d. Wissenschaften)

[16] Cotton E 1899 *Ann. Fac. Sci. Toulouse (II)* **1** 385

[17] York J W 1971 "Gravitational Degrees of Freedom and the Initial-Value Problem" *Phys. Rev. Lett.* **26** 1656

[18] Gödel K 1949 "An Example of a New Type of Cosmological Solution of Einstein's Field Equations of Gravitation" *Rev. Mod. Phys.* **21** 447

[19] Wainwright J 1979 "A Classification Scheme for Non-Rotating Inhomogeneous Cosmologies" *J. Phys. A: Math. Gen.* **12** 2015

[20] Collins C B and Stewart J M 1971 "Qualitative Cosmology" *Mon. Not. Roy. Astr. Soc.* **153** 419

[21] Hsu L and Wainwright J 1986 "Self-Similar Spatially Homogeneous Cosmologies: Orthogonal Perfect Fluid and Vacuum Solutions" *Class. Quantum Grav.* **3** 1105

[22] Ellis G F R and Bruni M 1989 "Covariant and Gauge-Invariant Approach to Cosmological Density Fluctuations" *Phys. Rev.* D **40** 1804

[23] Bruni M, Dunsby P K S and Ellis G F R 1992 "Cosmological Perturbations and the Physical Meaning of Gauge-Invariant Variables" *Astrophys. J.* **395** 34

[24] Lemaître G 1933 "L'Univers en Expansion" *Ann. Soc. Sci. Bruxelles I* **A 53** 51

[25] Tolman R C 1934 "Effect of Inhomogeneity in Cosmological Models" *Proc. Nat. Acad. Sci. U.S.* **20** 69

[26] Bondi H 1947 "Spherically Symmetric Models in General Relativity" *Mon. Not. Roy. Astr. Soc.* **107** 410

[27] Bruni M, Matarrese S and Pantano O 1995 "A Local View of the Observable Universe" *Phys. Rev. Lett.* **74** 1916

[28] Eisenhart L P 1933 *Continuous Groups of Transformations* (Princeton: Princeton University Press) Reprinted: 1961 (New York: Dover)

[29] Schouten J A 1954 *Ricci-Calculus* (Berlin: Springer-Verlag)

[30] Kantowski R and Sachs R K 1966 "Some Spatially Homogeneous Anisotropic Relativistic Cosmological Models" *J. Math. Phys.* **7** 443

[31] MacCallum M A H 1979 "Anisotropic and Inhomogeneous Relativistic Cosmologies" *General Relativity: An Einstein Centenary Survey* ed S W Hawking and W Israel (Cambridge: Cambridge University Press)

[32] Bonnor W B and Ellis G F R 1986 "Observational Homogeneity of the Universe" *Mon. Not. Roy. Astr. Soc.* **218** 605

[33] Matzner R, Rothman T and Ellis G F R 1986 "Conjecture on Isotope Production in the Bianchi Cosmologies" *Phys. Rev.* D **34** 2926

[34] Collins C B and Ellis G F R 1979 "Singularities in Bianchi Cosmologies" *Phys. Rep.* **56** 67

[35] Farnsworth D L 1967 "Some New General Relativistic Dust Metrics Possessing Isometries" *J. Math. Phys.* **8** 2315

* * *

Table 1: Survey of the existing LRS perfect fluid spacetime geometries with barotropic equation of state $p = p(\mu)$ [ $(\mu + p) > 0$ ]. References to subsections which discuss special subcases are given.

| | **LRS class I** | **LRS class II** | **LRS class III** |
|---|---|---|---|
| $(\omega/c)$ | $\neq 0$ | 0 | 0 |
| $k$ | 0 | 0 | $\neq 0$ |
| $(\Theta/c)$ | 0 | • $\neq 0$<br>• 0, static (7.1) | $\neq 0$ |
| $(\sigma/c)$ | 0 | • $\neq 0$<br>• 0 (6.2) | • $\neq 0$<br>• 0, ${}^3R > 0$ - FLRW (5.3) |
| $(\dot{u}/c^2)$ | • $\neq 0$<br>• 0 (4.2) | • $\neq 0$<br>• 0, dust/"silent" (6.1) | 0 |
| $a$ | • $\neq 0$<br>• 0, "unphysical" (4.1) | • $\neq 0$ (includes tilted SH (7.3))<br>• 0, "silent" *or* OSH/KS (6.3/7.2) | 0 |
| $E$ | • $\neq 0$<br>• 0, "magnetic" (4.4) | • $\neq 0$<br>• 0, FLRW (7.2.1) | • $\neq 0$<br>• 0, "magnetic" (5.2) |
| $H$ | • $\neq 0$<br>• 0 (4.3) | 0 | • $\neq 0$<br>• 0, ${}^3R > 0$ - FLRW (5.3) |